%% file: LUM-20-001_temp.tex
\documentclass[11pt,twoside,a4paper,cmspaper,final,collab]{cms-tdr}
\def\svnVersion{301847b-D}\def\svnDate{2026/06/15}\def\cmsCernNoTag{CERN-EP-2026-134}\def\cmsCernDate{\today}\def\cmsMessage{Submitted to Physical Review X}
\begin{document}\cmsNoteHeader{LUM-20-001}

\xspaceaddexceptions{\%\fbinv}
\newcommand{\pp}{\ensuremath{\Pp\Pp}\xspace}
\newcommand{\sigmaVis}{\ensuremath{\sigma_{\text{vis}}}\xspace}
\newcommand{\vdM}{vdM\xspace}
\newcommand{\betastar}{\ensuremath{\beta^{\ast}}\xspace}
\newcommand{\Sigx}{\ensuremath{\Sigma_x}\xspace}
\newcommand{\Sigy}{\ensuremath{\Sigma_y}\xspace}
\newcommand{\Delx}{\ensuremath{\Delta_x}\xspace}
\newcommand{\Dely}{\ensuremath{\Delta_y}\xspace}

\newcommand{\posU}[1]{\ensuremath{u_{\text{#1}}}\xspace}
\newcommand{\posV}[1]{\ensuremath{v_{\text{#1}}}\xspace}
\newcommand{\DelU}[1]{\ensuremath{\Delta\posU{#1}}\xspace}
\newcommand{\DelV}[1]{\ensuremath{\Delta\posV{#1}}\xspace}

\newcommand{\lsTrNom}{\ensuremath{\alpha_{\text{TR}/\text{NOM}}}\xspace}
\newcommand{\lsTrBpm}{\ensuremath{\alpha_{\text{TR}/\text{BPM}}}\xspace}
\newcommand{\lsBpmNom}{\ensuremath{\alpha_{\text{BPM}/\text{NOM}}}\xspace}

\newcommand{\sigZ}{\ensuremath{\sigma^{\PZ\to\PGm\PGm}}\xspace}
\newcommand{\sigZfid}{\ensuremath{\sigma^{\PZ\to\PGm\PGm}_\text{fid}}\xspace}
\newcommand{\effZmm}{\ensuremath{\epsilon^{\PZ\to\PGm\PGm}}\xspace}
\newcommand{\AZmm}{\ensuremath{A^{\PZ\to\PGm\PGm}}\xspace}
\newcommand{\lowPU}{\ensuremath{\text{low-PU}}\xspace}
\newcommand{\highPU}{\ensuremath{\text{high-PU}}\xspace}
\newcommand{\lumiZ}{\ensuremath{\mathcal{L}^\PZ_\highPU}\xspace}
\newcommand{\NZ}{\ensuremath{N^\PZ}\xspace}

\newcommand{\intlumi}{\lumi}
\newcommand{\instlumi}{\ensuremath{\lumi_\text{inst}}\xspace}

\providecommand{\cmsTable}[1]{\resizebox{\textwidth}{!}{#1}}

\newcommand{\NEG}{\makebox[0pt][r]{$-$}}

\cmsNoteHeader{LUM-20-001}

\title{Precision luminosity measurement in proton-proton collisions at a center-of-mass energy of \texorpdfstring{13\TeV}{13 TeV} with the CMS detector at the Large Hadron Collider}

\author[cern]{The CMS Collaboration}

\date{\today}

\abstract{
Discovering new fundamental physics requires spotting subtle deviations between theoretical predictions and experimental data. This delicate comparison hinges on the precise knowledge of the integrated luminosity, the measure of how many particle interactions were actually delivered by the collider. Here, we report a landmark measurement of the integrated luminosity by the Compact Muon Solenoid (CMS) experiment for proton-proton collisions at a center-of-mass energy of 13\TeV at the CERN Large Hadron Collider (LHC). By calibrating multiple independent monitors through specialized beam-separation techniques and rigorously validating their long-term stability against well-understood \PZ boson production rates, we comprehensively map and minimize systematic uncertainties. Combining the findings yields a total integrated luminosity precision of 0.73\% for the entire data set. This marks the most precise luminosity measurement ever achieved at a bunched-beam hadron collider. Crossing the sub-percent precision threshold per data taking year fundamentally sharpens our ability to test the standard model and establishes a vital baseline for the upcoming High-Luminosity LHC era.
}

\hypersetup{
pdfauthor={CMS Collaboration},%
pdftitle={Precision luminosity measurement in proton-proton collisions at sqrt(s)=13 TeV with the CMS detector},%
pdfsubject={CMS},%
pdfkeywords={CMS, LHC, luminosity, vdM, calibration}}

\maketitle

\section{Introduction}

Instantaneous luminosity (\instlumi) is one of the primary quantities that characterize the performance of hadron colliders, such as the CERN Large Hadron Collider (LHC). It is governed by the properties of the colliding beams and quantifies the interaction rate as two particle bunches cross. Integrating \instlumi over time and over all colliding bunch pairs yields the integrated luminosity, hereafter referred to simply as luminosity (\intlumi). 
This quantity determines the total number of events for a given process and therefore represents the size of the data set delivered to an experiment. 
Luminosity provides the normalization for cross section measurements,  making its precise determination essential. Its uncertainty is a key contributor to the overall uncertainty in many collider physics analyses~\cite{LUMI_Motiv_Mangano}, and it is the dominant systematic uncertainty in several important measurements performed by the CMS experiment at the CERN LHC~\cite{Hayrapetyan:2898867, CMS:2018fks, CMS:2019raw, CMS:2020cph, CMS:2023qyl, CMS:2024myi}.
While the original design goal for the 
systematic uncertainty at CMS was 5\%~\cite{CERN/LHCC-2006-001}, with the LHC evolving into a precision instrument, increasingly stringent accuracy requirements emerged and will continue to do so in future running conditions.
In the High-Luminosity LHC era, precise luminosity determination will become more challenging due to the high-pileup conditions~\cite{ATLAS:2019mfr,Collaboration:2759074}, \ie, due to the large number of simultaneous proton-proton interactions. Achieving sub-percent uncertainty nevertheless remains essential to fully
unlock the broad physics reach of the LHC and of the CMS experiment~\cite{Collaboration:2706512, Collaboration:2759074}.

At the LHC, two oppositely circulating bunched proton (\Pp) or ion beams collide at four interaction points (IPs) corresponding to the ALICE, ATLAS, CMS, and LHCb experiments. 
The LHC orbit is divided into 25\unit{ns} time intervals, resulting in 3564 bunch slots per beam. Each slot may be filled with colliding bunch pairs, contain a bunch in only one beam, or remain empty. These slots are labeled by bunch crossing identification numbers (BCIDs), ranging from 1 to 3564.

In principle, the \instlumi can be determined from the beam parameters, in particular from the proton density distributions of the bunches.
In practice, however, these quantities cannot be measured with sufficient precision under standard physics data-taking conditions.
Instead, luminosity is typically inferred from the measured rate ($R$) of 
a well-understood and precisely measured process
by scaling the rate with the corresponding calibration factor, commonly referred to as the visible cross section (\sigmaVis)~\cite{Aaij:2014ida, CMS:2021xjt, ALICE:2022xir, ATLAS:2022hro}:
\begin{equation}
    \instlumi(t)=\frac{R(t)}{\sigmaVis}.
\end{equation}
The value of \sigmaVis is obtained using the van der Meer (\vdM) calibration method~\cite{Grafstrom:2015foa}, described in Section~\ref{sec:VdMCalib}, based on transverse beam separation scans~\cite{VdM:1968, CMS:2021xjt}, performed under special data-taking conditions.  Such scans are typically performed once per year for a given beam type, such as protons or different kinds of ions.
At the CMS experiment, $R$ may correspond to a range of quantities independently measured by dedicated detectors, referred to as ``luminometers'' (Section~\ref{sec:the_cms_detector}), which are operated and maintained by the CMS Beam Radiation, Instrumentation and Luminosity (BRIL) project~\cite{Guthoff:2209346, CMS:2023gfb}.
Alternatively, $R$ represents the rate of a well-understood physics process,
such as \PZ boson production with subsequent decay to a pair of oppositely charged muons.
Measurements based on such processes facilitate cross-checks of the \sigmaVis calibrations across different years, as well as the long-term stability of the luminosity measurement~\cite{Salfeld-Nebgen:2018bdp, CMS:2024zco}.

Luminosity measurements based on the \vdM method have been reported by all four LHC experiments for the data sets collected in the 2015--2018 LHC running period (Run~2). 
For \pp collisions at a center-of-mass energy ($\sqrt{s}$) of 13\TeV, the CMS Collaboration measured the integrated luminosity of the combined 2015--2016 data set with an uncertainty of 1.2\%~\cite{CMS:2021xjt}, while the ATLAS Collaboration obtained an uncertainty of 0.83\% for the combined 2015--2018 data set~\cite{ATLAS:2022hro}.
For lead-lead collisions recorded at a center-of-mass energy per nucleon pair 
of 5.02\TeV in 2015 and 2018, the ALICE and CMS Collaborations each reported combined luminosity measurements with uncertainties of 2.5 and 1.6\%, respectively~\cite{ALICE:2022xir, LUM-20-002}.

In this paper, we report the measurement of the integrated luminosity for the \pp collision data set at $\sqrt{s}=13\TeV$ recorded by the CMS experiment in 2017--2018.
We highlight new techniques that enable 
sub-percent precision
by constraining the key effects that can bias the measured \sigmaVis values or impact the calibration stability over time and under changing running conditions.  
In particular, significant improvements in precision relative to the preliminary results reported in~\cite{CMS-PAS-LUM-17-004} and~\cite{CMS-PAS-LUM-18-002} are achieved through a better understanding of beam-beam effects and their uncertainties, as well as of systematic and random orbit drifts. Further gains arise from improved treatments of nonfactorization of the proton density in transverse coordinates and the calibration of the LHC transverse length scale (LS), as well as refined detector corrections, and the averaging of independently calibrated luminometers.
We also report the combination of this measurement with the 2015--2016 result, which provides an overall uncertainty for the full Run~2 \pp CMS data set.

The paper is structured as follows:
Section~\ref{sec:the_cms_detector} provides a description of the CMS detector and its luminometers. Section~\ref{sec:VdMCalib} describes the \vdM method for luminosity calibration, along with the various effects that determine the accuracy of \sigmaVis and the techniques used to constrain them. Section~\ref{sec:detector_corrections} presents the detector-specific effects that impact the measurement of $R$. Section~\ref{sec:luminosity_integration_and_uncertainty} discusses the uncertainties associated with high-luminosity running and varying data-taking conditions.
Finally, the most precise results achieved so far at bunched hadron colliders are reported in Section~\ref{sec:results}, and a summary is given in Section~\ref{sec:summary}.

\section{The CMS detector and its luminometers}
\label{sec:the_cms_detector}

The CMS apparatus~\cite{CMS:2008xjf, CMS:2023gfb} is a multipurpose, nearly hermetic detector, designed to trigger on~\cite{CMS:2020cmk, Khachatryan:2016bia, CMS:TRG-19-001} and identify electrons, muons, photons, and (charged and neutral) hadrons~\cite{CMS:2020uim, MUO-16-001, TRK-11-001}.
Its central feature is a superconducting solenoid of 6\unit{m} internal diameter, providing a magnetic field of 3.8\unit{T}.
Within the solenoid volume are a silicon pixel and strip tracker for measuring charged particles in the pseudorapidity range of $\abs{\eta}<2.5$; a lead tungstate crystal electromagnetic calorimeter for measuring the energy of photons and electrons; and a brass and scintillator hadron calorimeter for measuring the energy of charged and neutral hadrons.
Each system is composed of a barrel and two endcap sections.
The hadron forward (HF) calorimeter, using steel as an absorber and quartz fibers as the sensitive material, extends the $\eta$ coverage provided by the barrel and endcap detectors.
The two halves of the HF are located 11.2\unit{m} from the IP, one on each end, and together they provide coverage in the range $3.0<\abs{\eta}<5.2$~\cite{HCAL:2007}.
Muons are reconstructed in the range $\abs{\eta}<2.4$ using gas-ionization detectors embedded in the steel flux-return yoke outside the solenoid, with detection planes made using three technologies: drift tubes, cathode strip chambers, and resistive-plate chambers~\cite{MUO-16-001}.
More detailed descriptions of the CMS detector, together with a definition of the coordinate system used and the relevant kinematic variables, can be found in Refs.~\cite{CMS:2008xjf,CMS:2023gfb}.

Events of interest are selected using a two-tiered trigger system.
The first level, composed of custom hardware processors, uses information from the calorimeters and muon detectors to select events at a rate of around 100\unit{kHz} within a fixed latency of 4\mus~\cite{CMS:2020cmk}.
The second level, known as the high-level trigger, consists of a farm of processors running a version of the full event reconstruction software optimized for fast processing, and reduces the event rate to a few kHz before data storage~\cite{Khachatryan:2016bia, CMS:TRG-19-001}.

The CMS experiment employs multiple luminometers throughout the \vdM calibration and regular physics data-taking periods to ensure a redundant and consistent luminosity determination. Detectors originally designed for other purposes as well as dedicated instruments are leveraged for this objective.
A detailed overview of the luminometers and algorithms used to estimate \instlumi during 2015--2016, as well as a depiction of the locations of these subsystems within CMS, is provided in Ref.~\cite{CMS:2021xjt}.
Here, we briefly introduce the luminometers employed in 2017--2018 and highlight relevant changes with respect to 2015--2016.

Three luminometers provide online bunch-by-bunch luminosity measurements via a dedicated data acquisition (DAQ) system that operates independently of the central CMS DAQ~\cite{Khachatryan:2016bia,Tapper:1556311}. This independent DAQ system is critical for optimizing LHC operations and real-time data collection.

Luminometry based on the HF employs a dedicated readout and uses a restricted region ($3.15 < \abs{\eta} < 3.5$) to ensure uniform occupancy. Two luminosity algorithms are implemented for HF. 
The occupancy method (HFOC) calculates the average number of hits over a threshold per tower. The transverse energy sum method (HFET) ---introduced in 2017--- computes the scalar sum of transverse energy deposited in the selected HF towers.
The HFET observable shows improved intrinsic linearity and a reduced magnitude of data-driven corrections, as 
detailed in Section~\ref{sec:detector_corrections}.

The pixel luminosity telescope (PLT) consists of 16 silicon-pixel ``telescopes''~\cite{CMSBRIL:2022qin} arranged around the beam pipe 
approximately 1.7\unit{m} from the interaction point 
at $\abs{\eta}\approx 4.2$. 
Each telescope is a stack of three pixel sensor planes that form a tracklet by requiring coincident hits in all planes. This approach identifies charged particles originating from the interaction point while suppressing activation hits and noise. 

The beam conditions monitor with fast readout (BCM1F), housing 10 silicon, 10 polycrystalline diamond, and 4 single-crystal diamond sensors~\cite{Guthoff:2019cil}, was used as a real-time luminometer and beam-induced background monitor. However, it is not included in the integrated luminosity determination presented here as radiation damage significantly degraded its performance, especially toward the end of 2018. The BCM1F and PLT detectors were upgraded in the beginning of 2017.

The rate measurements for HFOC, PLT, and BCM1F use the zero-counting method~\cite{CMS:2021xjt}, where the probability of zero hits is used to infer the expected number of hits based on the fact that hit counts follow a Poisson distribution. The transverse energy sum measured in HFET is proportional to the luminosity. 

A bunch-by-bunch offline luminosity measurement is provided by the pixel cluster counting (PCC) method, which uses the average number of clusters measured in the silicon pixel tracker per unbiased event. In regular physics data-taking conditions, two PCC data streams are recorded via the CMS DAQ: ``zero-bias-triggered'' data for colliding BCIDs (2\unit{kHz} bandwidth) for luminosity counting and ``random-triggered'' data for all BCIDs (500\unit{Hz} bandwidth) for data-driven correction for out-of-time hits. In the vdM calibration data-taking periods, the zero-bias triggers were gated on five colliding BCIDs and recorded events with a total rate of 20\unit{kHz}, thereby ensuring a high event count even at large beam separations.

Additionally, two low-occupancy systems provide orbit-integrated luminosity measurements. The DT method utilizes the drift tubes and resistive plate chambers of the barrel muon system to count muon track trigger primitives reconstructed in the barrel muon track finder.
The radiation monitoring system for the environment and safety (RAMSES) measures the ambient dose equivalent rate with ionization chambers located in the experimental cavern~\cite{RAMSES2005, Ledeul:PCaPAC2018-FRCC3}. 
Several such detectors are located in the CMS cavern. Two of them, positioned on the HF platforms on either side of CMS, are utilized for luminosity measurement.

\section{Absolute luminosity calibration with the \texorpdfstring{\vdM}{vdM} method}
\label{sec:VdMCalib}

\begin{figure*}[!ht]
\centering
\includegraphics[width=0.45\textwidth]{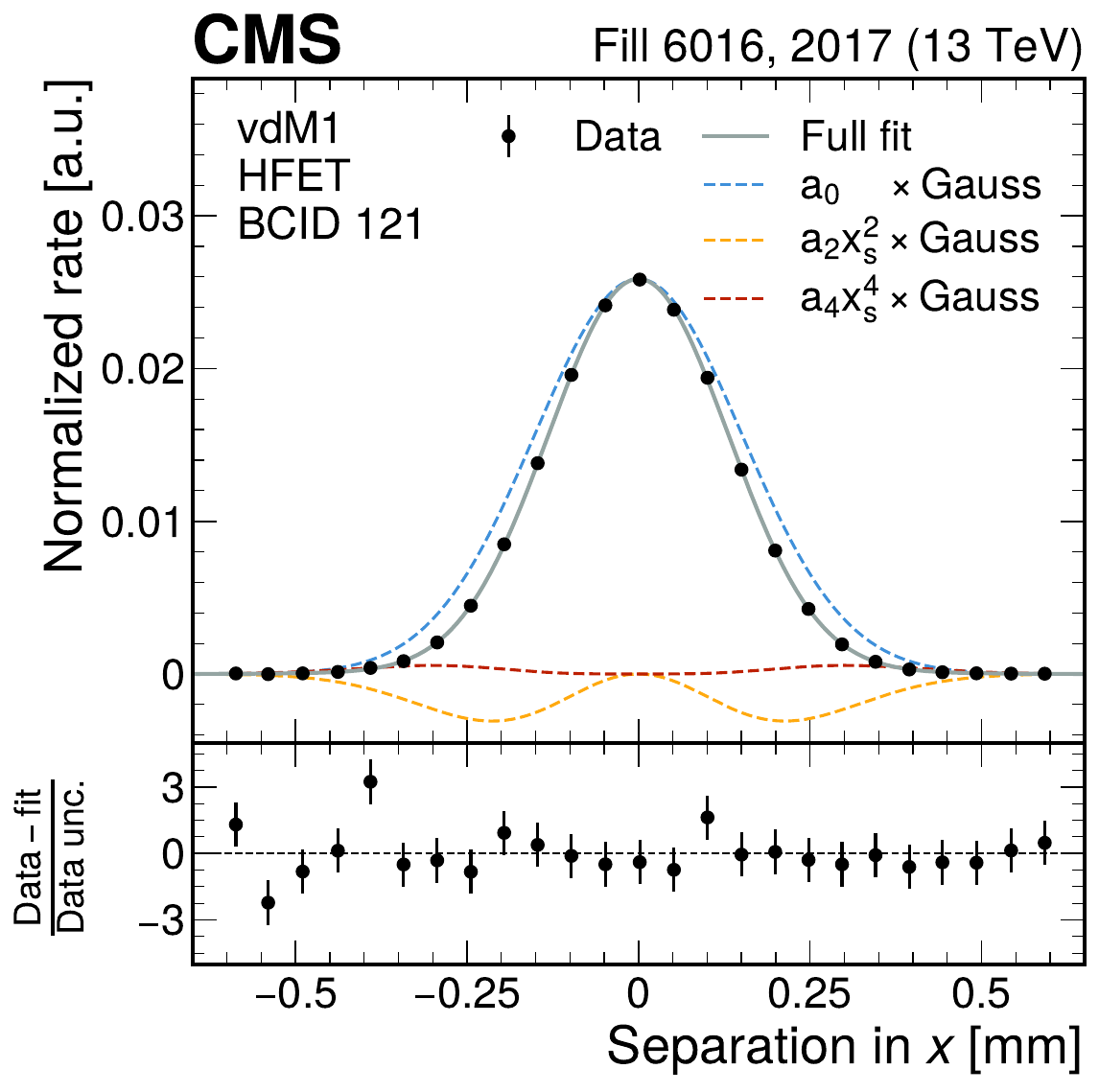}
\hspace{0.05\textwidth}
\includegraphics[width=0.45\textwidth]{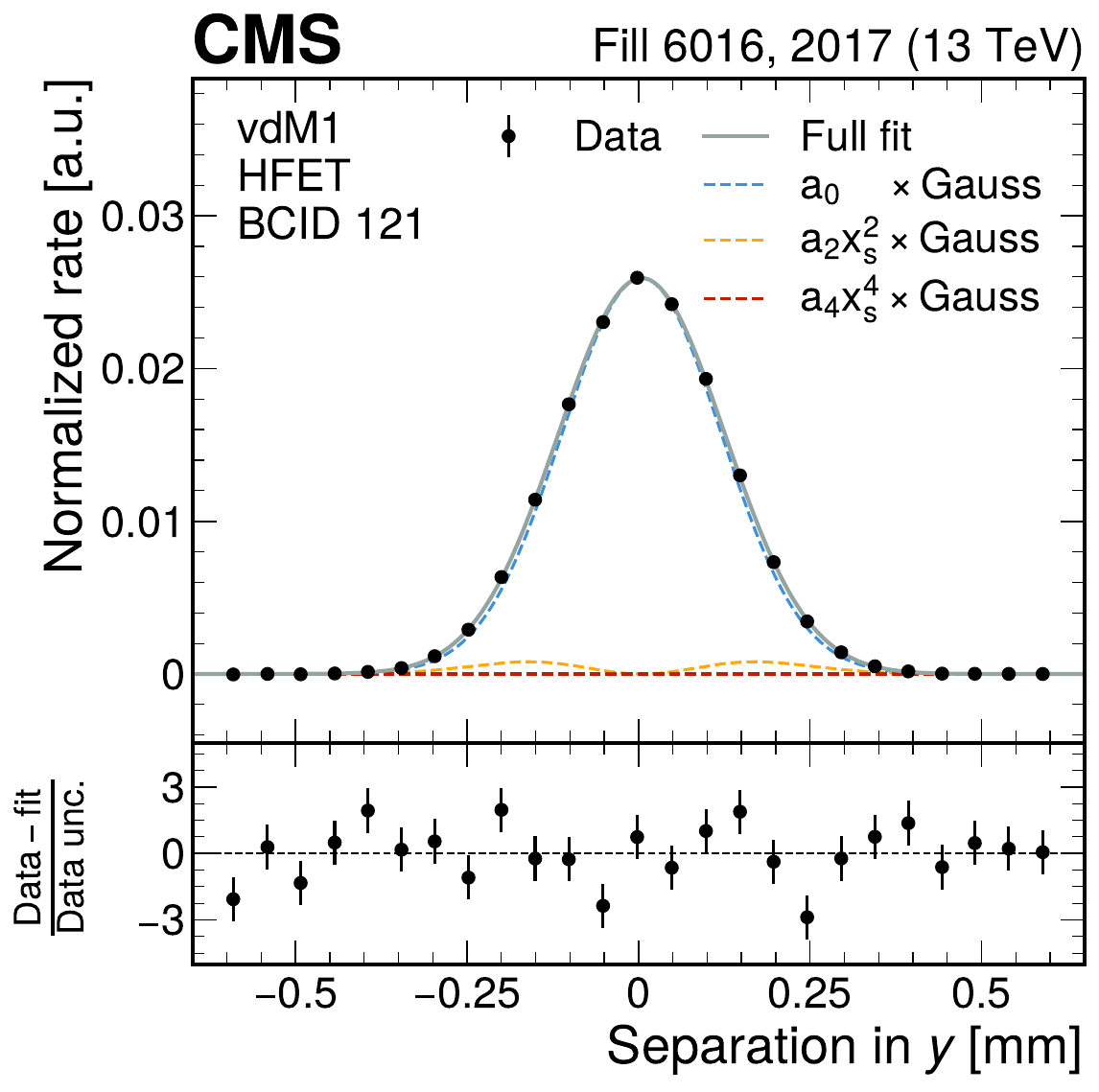}
\caption{
    An illustration of the \vdM fit procedure for the HFET method in the 2017 \vdM fill after all corrections are applied.
    The luminometer rate (normalized by the product of the bunch proton multiplicity values) is shown for a single bunch (BCID 121) as a function of the transverse separation of the two beams in the horizontal (left) and vertical (right) directions.
    The data are fitted by the product of a fourth-order positive symmetric polynomial and a Gaussian function.
    The three additive components, as well as the full analytic function, are plotted. The variable $x_s$ signifies coordinates standardized with the width and mean of the Gaussian function, $x_s=(x-\mu)/\sigma$.
    The effective widths of the fitted curves, \Sigx and \Sigy, are defined by Eq.~\eqref{eq:capsigma}, and the peak value is the head-on rate normalized by the bunch intensities ($R(0,0)/(N_1N_2)$) in Eq.~\eqref{eq:SigmaVis}.
    The lower panel shows the difference between the measured data and the fitted function, divided by the statistical uncertainty of the normalized rate.
}
\label{fig:vdMcurve}
\end{figure*}

Pairs of transverse beam separation scans, called \vdM scans, are analyzed to determine \sigmaVis, which depends both on the detector used and the counting method employed.
During a \vdM scan, the two beams are separated in several steps in one transverse direction ($x$, which is horizontal or $y$, which is vertical) and the luminometer rate ($R$) is measured as a function of the distance between the two beam orbits at the interaction point (beam separation \Delx and \Dely).
An example of the data recorded during an $x$-$y$ \vdM scan
is shown in Fig.~\ref{fig:vdMcurve}.

In the \vdM method, the transverse sizes of the convolved bunches \Sigx and \Sigy, the so-called bunch convolution widths, are determined as in Ref.~\cite{Grafstrom:2015foa}:
\ifthenelse{\boolean{cms@external}}{
\begin{equation}\begin{aligned}\label{eq:capsigma}
    \Sigx&=\frac{1}{\sqrt{2\pi}R(0,0)}\int\!R(\Delx,0)\,\rd\Delx,\\
    \Sigy&=\frac{1}{\sqrt{2\pi}R(0,0)}\int\!R(0,\Dely)\,\rd\Dely,
\end{aligned}\end{equation}
}{
\begin{equation}\label{eq:capsigma}
    \Sigx=\frac{1}{\sqrt{2\pi}R(0,0)}\int\!R(\Delx,0)\,\rd\Delx,
    \quad
    \Sigy=\frac{1}{\sqrt{2\pi}R(0,0)}\int\!R(0,\Dely)\,\rd\Dely,
\end{equation}
}
where the measured rate ($R$) is considered as a two-dimensional (2D) function of \Delx and \Dely, such that $R(\Delx,0)$ and $R(0,\Dely)$ are the rates measured during the \vdM scans in $x$ and $y$, respectively, and $R(0,0)$ is the rate corresponding to head-on collisions.
In the case of a Gaussian luminosity profile \Sigx and \Sigy are simply the widths of the Gaussian function. 
When the transverse bunch proton density ($\rho(x,y)$) can be factorized into the product of two 1D functions, 
\begin{equation}\label{eq:fact}
    \rho(x,y)=\rho_x(x)\rho_y(y), 
\end{equation}
then $R$ also has this property, and the visible cross section, a double integral of $R$, can be calculated based the parameters obtained from the $x$ and $y$ scans:
\ifthenelse{\boolean{cms@external}}{
\begin{equation}\begin{aligned}\label{eq:SigmaVis}
    \sigmaVis&=\frac{1}{N_1N_2f_\text{LHC}}\iint\!R(\Delx,\Dely)\,\rd\Delx\rd\Dely \\
    &= 2\pi\frac{R(0,0)}{N_1N_2f_\text{LHC}}\Sigx\Sigy,
\end{aligned}\end{equation}
}{
\begin{equation}\label{eq:SigmaVis}
    \sigmaVis=\frac{1}{N_1N_2f_\text{LHC}}\iint\!R(\Delx,\Dely)\,\rd\Delx\rd\Dely = 2\pi\frac{R(0,0)}{N_1N_2f_\text{LHC}}\Sigx\Sigy,
\end{equation}
}
where $N_1$, $N_2$ are the numbers of protons in the two colliding bunches and $f_\text{LHC}\approx11245$\unit{Hz} is the LHC orbit frequency. 
In practice, the data measured during a \vdM scan are fitted by a continuous function, as shown in Fig.~\ref{fig:vdMcurve}, and this function is then used to calculate \Sigx or \Sigy according to Eq.~\eqref{eq:capsigma}.
In this paper, a Gaussian function multiplied by a positive, symmetric fourth-order polynomial (``Poly4G'') is used for the fit of the \vdM scan curves.

\begin{figure*}[!htbp]
\centering
\includegraphics[width=0.92\textwidth]{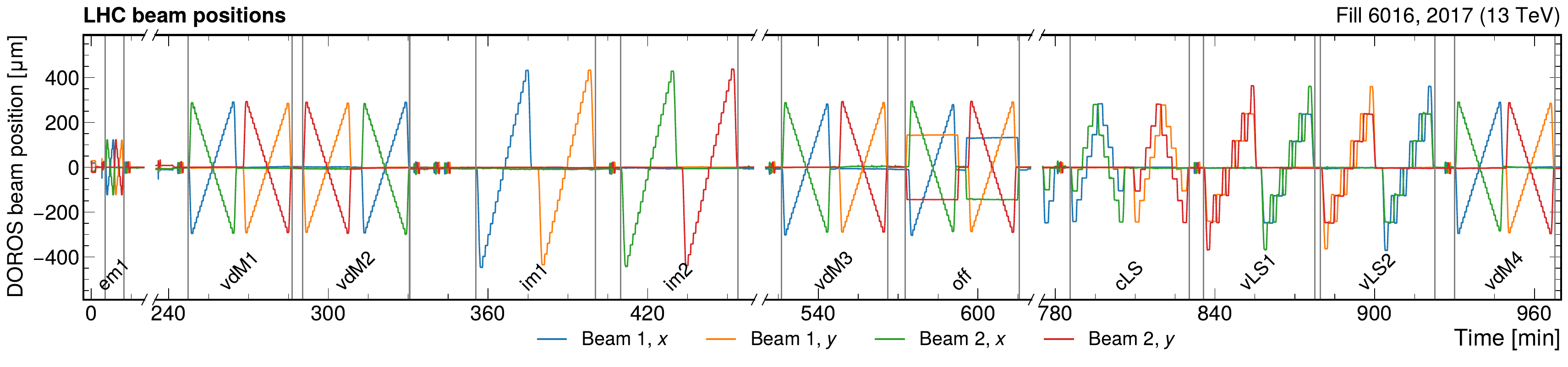} \\
\includegraphics[width=0.92\textwidth]{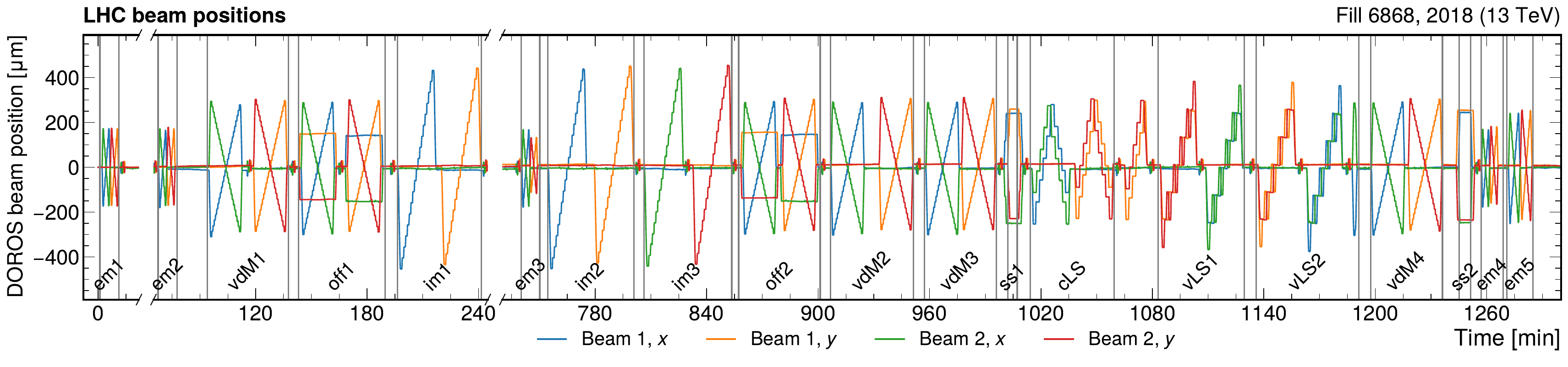}
\caption{
    Beam positions in $x$ and $y$ for both LHC beams as functions of time during LHC fills 6016 (upper) and 6868 (lower), as measured with the DOROS BPMs with 1\unit{s} time granularity.
    Time is indicated relative to the first included data point.
    The beginning and end of individual scans are indicated by vertical lines.
    Each \vdM-like scan pair consists of 
    a scan along the $x$ axis and one along the $y$ axis,
    and is labeled by the abbreviation of the corresponding scan type: ``vdM'', ``em'', ``im'', and ``off'' for standard \vdM, emittance, beam imaging, and offset scans, respectively.
    Variable- and constant-separation length scale scans are marked with ``vLS'' and ``cLS'', respectively.
    The super separation periods in 2018 are marked with ``ss''.
}
\label{figures:orbitdrift:dorosvstime}
\end{figure*}

The \vdM calibration programs for the 2017 and 2018 data-taking periods were performed during the LHC fills 6016 and 6868, respectively.
The beam orbits at the IP during these two fills were monitored with beam position monitor (BPM) systems operated by the LHC control room. The diode orbit and oscillation (DOROS) BPMs~\cite{Gasior:2313935, Gasior:1476070, doros} are located 21.6\unit{m} away from the IP on either side and directly measure the beam displacements during the \vdM scans.
The arc BPMs~\cite{CERN-ACC-NOTE-2013-0006}, located in the LHC arcs \mbox{L 33-9} and \mbox{R 9-33}, are outside of the steering magnets and are only indirectly sensitive to the beam displacements. They measure the deviation from the reference orbit.
To determine the beam positions, the readings are extrapolated to the IP using a model of the LHC optics. 
The comparison of the measurements from the two independent systems allows for an estimation of the uncertainty associated with BPM 
instrumentation effects.
The beam positions measured by the DOROS BPMs during the two programs are shown in Fig.~\ref{figures:orbitdrift:dorosvstime}.
In addition to standard \vdM scans, these programs also included emittance (em), beam imaging (im), and offset (off) scans, as well as variable- and constant-separation length scale (vLS and cLS) scans and super separation (ss) periods.
The different scan types are used to constrain systematic uncertainties in the estimation of \sigmaVis, as discussed in the following sections.

\subsection{Luminometer rates}

Background contributions to the luminometer rates measured during \vdM scans arise from detector noise and beam-induced background, such as beam-gas collisions.
To obtain the luminometer rate that corresponds to the nominal \pp collisions, the background rates must be subtracted.
In 2017 data, the background rates are estimated from the measured luminometer rates during the \vdM scan steps with the largest beam separation, which reaches to more than 5.5 times the nominal bunch width in both transverse directions.
For 2018, the background rates are instead estimated from the luminometer rate measured in the super separation periods, during which the beams are kept separated by about 5 times the nominal bunch width simultaneously in each transverse direction.
A comparison between the 2018 \sigmaVis estimate derived using the super separation periods and the result obtained by applying the 2017 method to the 2018 \vdM scans shows agreement within 0.03\%.
For each luminometer, the estimated background rate is at most 1\% of the peak rate during \vdM scans. This value is subtracted from the measured rates before the fits for the extraction of \Sigx and \Sigy are performed.

To estimate potential systematic effects, an alternative background correction method is studied. In this approach, the background is determined independently for each scan step in order to account for the possible dependencies on the \instlumi and on time.
The combined contribution from detector noise and from interactions of a single bunch with the residual gas in the beam pipe ($NG$) is measured in bunch slots where only one beam contains protons nominally. 
The detector noise component ($N$) is measured separately at the end of the orbit, where no bunches are present for more than one hundred consecutive bunch slots.
The background, modeled as the sum of contributions from beam-gas interactions for each colliding bunch and from detector noise, is estimated as $2NG - N$.
Applying this correction to the \vdM rates on a per-step basis results in a -0.03\% (0.07\%) shift in 2017 (2018). 
An uncorrelated uncertainty of 0.07\% is assigned to both years. 

\subsection{Bunch intensity}

The number of protons in the colliding bunches, \ie, the bunch intensity, is evaluated using measurements by two LHC instruments.
The fast beam current transformer (FBCT)~\cite{fbct, fbct1, fbct2} is a toroidal transformer surrounding the LHC beamline with a sampling rate much higher than the LHC collision frequency; it can resolve the intensity of individual bunches.
The direct-current beam current transformer (DCCT)~\cite{Barschel:1425904} is also a toroidal transformer, but it employs a zero-flux feedback loop to measure the total beam current with a high precision of 0.2\%.

In contrast to the FBCT, the DCCT is insensitive to the individual bunch structure and is therefore only suitable for measuring the total beam intensity.
To achieve optimal precision for the bunch-by-bunch proton population, the sum of the FBCT measurements is normalized to the DCCT measurement.
An uncertainty of 0.2\% is assigned to the bunch intensity to account for the DCCT precision. Any uncertainty in the FBCT measurement propagates directly to the per-bunch visible cross section and is covered by the bunch-by-bunch consistency uncertainty discussed in Section~\ref{sec:sigmaVis}.

Spurious charge can be present in the accelerator that contributes to the FBCT and DCCT intensity measurements but not to collisions.
This charge is either located in nominally empty BCIDs (``ghost'' charge) or
is present in filled BCIDs but outside of the 2.5\unit{ns} time window containing the colliding particles (``satellite'' charge).
Ghost charge contributes to the DCCT measurement but not to the FBCT measurement of nominally filled BCIDs, while satellite charge contributes to both.

Both ghost and satellite contributions are measured using the LHC beam synchrotron radiation longitudinal density monitor (LDM)~\cite{Jeff:2012zz, Jeff:1513180} and are subtracted from the bunch intensities, as described in Ref.~\cite{CMS:2021xjt}.
The ghost fraction is also measured independently using the beam-gas imaging method~\cite{Aaij:2014ida, Barschel:1693671, Coombs:2019nmc}, which compares the beam-gas interaction rates in bunch crossings at the LHCb experiment where only one or neither bunch slot is filled. 
The difference between the LHCb and the LDM estimate is used as the ghost fraction uncertainty. 
Since no fully independent measurement is available for satellite charge, the uncertainty is estimated 
by omitting the baseline correction in the LDM measurement. 
This correction accounts for the signal observed at the edges of the radio-frequency bins. The resulting difference is assigned as the satellite uncertainty.

Applying the ghost charge corrections changes \sigmaVis by $(0.07\pm0.01)\%$ and $(0.28\pm0.04)\%$ in 2017 and 2018, respectively, while the satellite charge correction leads to shifts of 
$(0.02\pm0.05)\%$ and $(0.09\pm0.03)\%$.
Overall, applying the ghost and satellite corrections result in \sigmaVis changes of $(0.07\pm0.01)\%$ in 2017 and $(0.28\pm0.04)\%$ in 2018.
No significant time dependence of the total spurious charge correction is observed in 2017, whereas a small decrease of 0.15\% over time is seen in 2018.
To account for the possible correlation between the ghost and satellite charge estimation methods, the uncertainties are added linearly, yielding a total spurious charge uncertainty of 0.06\% in 2017 and 0.07\% in 2018, fully correlated between the two years.

\subsection{Beam-beam interaction}
\label{sec:BB}

The colliding bunches interact via their electromagnetic fields, resulting in a deflection of their orbits away from each other (beam-beam deflection), as well as in a distortion of the bunch proton densities (dynamic-$\beta$ effect)~\cite{BB_paper24}. 
The impact of these contributions is assessed separately for every colliding bunch pair and each vdM scan step.
The two effects bias the \sigmaVis estimation in opposite directions, resulting in partial cancellation. In addition, they share common systematic sources that induce anticorrelated shifts in their impact. To account for this, a single combined uncertainty is assigned to the two effects.

The beam-beam deflection, also called coherent shift, originates from a mutual repulsive angular kick between two bunches that collide with a nonzero transverse separation. 
This increases the nominal beam separation and reduces the luminosity.
The magnitude of the deflection can be computed analytically using the Bassetti--Erskine formula~\cite{Bass1,Babaev1}.

Under the Run~2 LHC conditions, the dynamic-$\beta$ effect (also called the incoherent or optical effect) generally leads to a squeezed bunch size, thereby increasing the \instlumi.
In addition, it can also perturb the internal structure of the bunches, leading to separation-dependent nonfactorization effects (i.e., a breakdown of Eq.~\eqref{eq:fact}). The applied corrections account for both effects.

The magnitude depends on the betatron tunes $Q_{x,y}$ that describe the frequency of the oscillations along the LHC orbit in the transverse directions, as well as on the beam-beam parameter of the colliding bunches that can be calculated from the transverse bunch size, the bunch intensity, the Lorentz factor of the particles, and the value of the $\beta$ function at the IP ($\betastar=19.17\pm1.92\unit{m}$ in the 2017--2018 \vdM fills~\cite{Wenninger:2668326, BB_paper24}).
In practical terms, the dynamic-$\beta$ effect is estimated using multi-particle simulations~\cite{Balagura:2020fuo, BB_paper24}.
For computational efficiency, the simulation code \textsc{b*b}~\cite{Balagura:2020fuo} based on a weak-strong model (where the transverse proton density of one bunch is deformed by the field of the other bunch that is considered unperturbed) is used to extract a 3D parametrization as a function of the betatron tunes and the beam-beam parameter in the limit of round bunches with an initially Gaussian distribution and of equal intensity colliding at a single IP with zero crossing angle~\cite{BB_paper24}.
Differences with respect to this ideal configuration are studied with \textsc{b*b} and with \textsc{combi}~\cite{Furuseth:2688401}, which uses a strong-strong model where perturbations of both bunches are considered.

The LHC operates at nominal betatron tune values of $Q_x=64.31\pm0.02$ and $Q_y=59.32\pm0.02$~\cite{Wenninger:2668326,BB_paper24}.
These tunes are perturbed during collisions with magnitude depending on the strength of the interaction, as well as the number of IPs where the bunches collide.
In the 2017 \vdM fill, the 32 bunch pairs colliding at the CMS IP also collided at the ATLAS IP but not in the other two experiments.
In 2018, a larger number of 124 bunches per beam was used, with additional collisions taking place for some of the bunches at the ALICE and/or LHCb IPs.
While about half of the colliding bunch pairs have collisions only at two IPs, five additional multicollision configurations have to be considered for the remaining colliding bunch pairs.
The multicollision bias can be taken into account by using a shifted tune for each configuration in the calculation of the dynamic-$\beta$ effect~\cite{BB_paper24}.

\begin{figure*}[!htpb]
\centering
\includegraphics[width=0.455\textwidth]{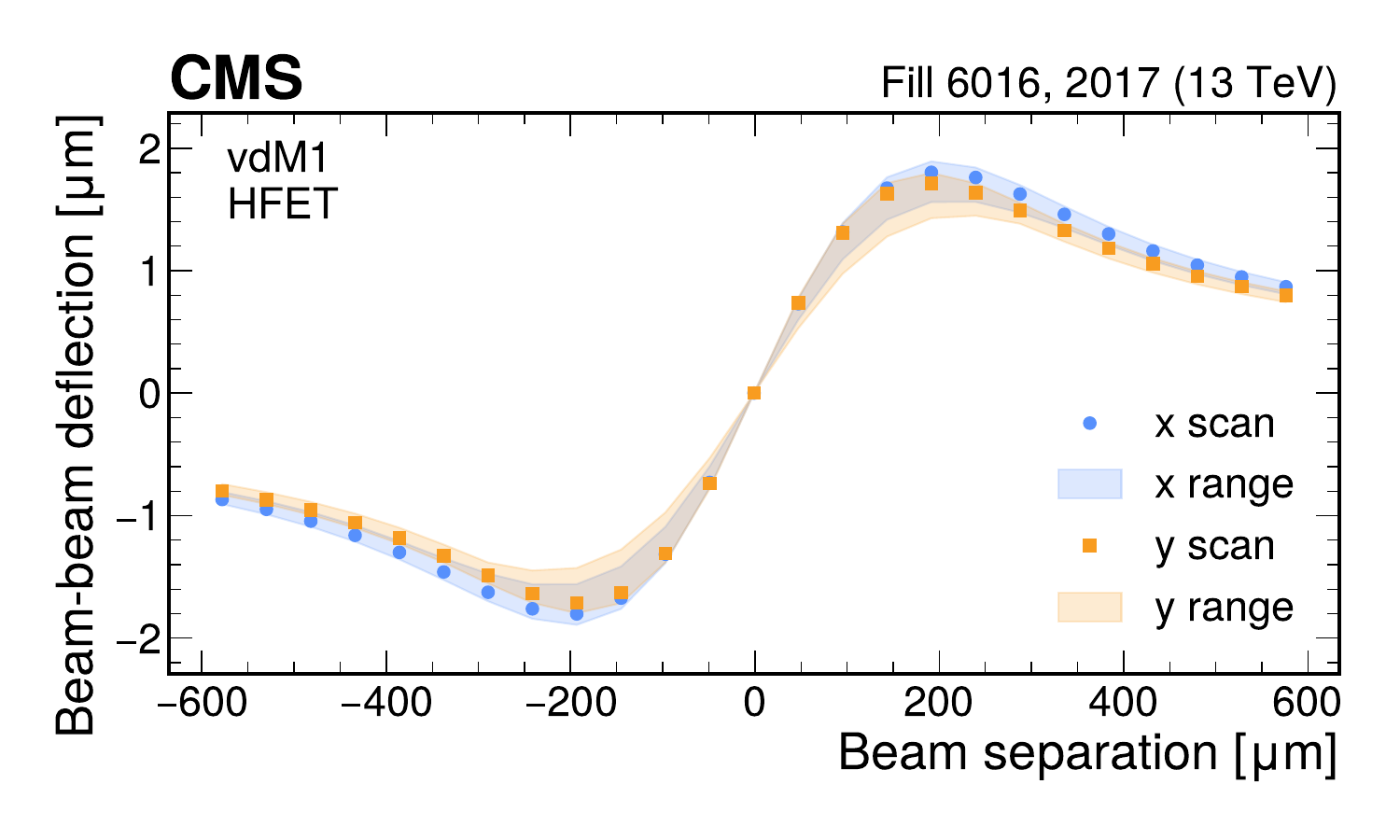}
\includegraphics[width=0.455\textwidth]{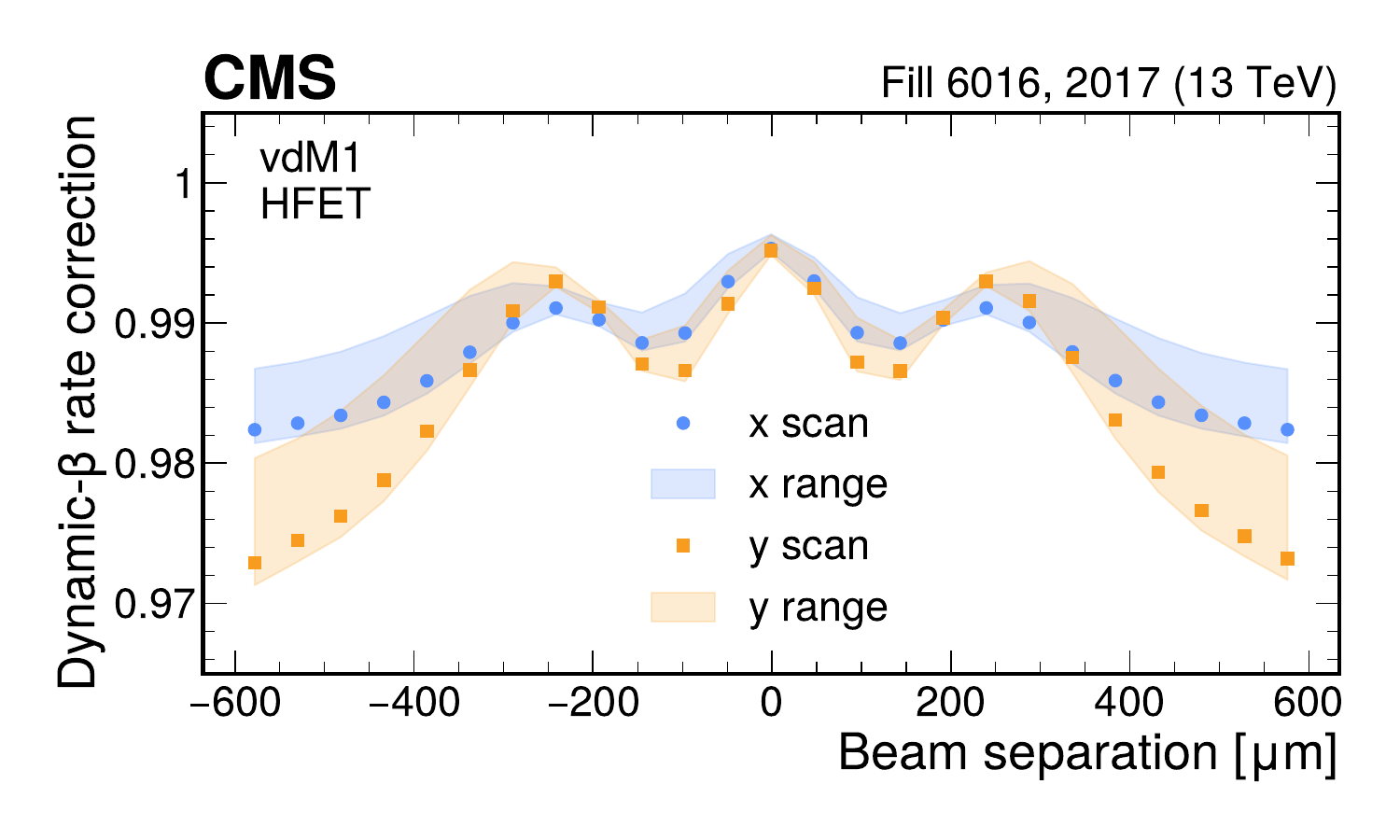}
\caption{
    Beam-beam deflection (left) and dynamic-$\beta$ correction (right) for the first scan pair in fill 6016, separately for scans in $x$ and $y$.
    The corrections are computed individually for each colliding bunch pair.
    The points represent the average over all bunch pairs, and the shaded area covers the minimum and maximum values obtained for any bunch pair.
}
\label{fig:bb:corr}
\end{figure*}

Figure~\ref{fig:bb:corr} shows the corrections for the beam-beam effects for a representative 2017 \vdM scan pair.
The average impact on \sigmaVis from beam-beam deflection and the dynamic-$\beta$ effect is $+1.71$ and $-1.08\%$ in 2017, and $+1.61$ and $-1.01\%$ in 2018, respectively.
The dynamic-$\beta$ effect shows a time dependence driven by the increase of the round-beam-equivalent beam-beam parameter during the fills.

The leading uncertainties in the beam-beam interaction model are the knowledge of the nominal tunes (0.18\% in 2017 and 0.19\% in 2028), the value of \betastar (0.07\% and 0.08\%), and the phase advances between the IPs in the simulation (0.04\% and 0.10\%)~\cite{BB_paper24}.
Additional contributions stem from nonideal beam properties, particularly the observed non-Gaussian tails (0.15\% and 0.16\%), the ellipticity of the bunch shapes (0.03\% in both years), and the imbalance in terms of transverse bunch sizes between the two beams (0.10\% and 0.05\%).
The parameterization of the \textsc{b*b} results introduces an additional uncertainty of 0.1\%, while other effects have a negligible impact ($<$0.01\%).
Adding these sources in quadrature, the total uncertainties are 0.29 and 0.30\% in 2017 and 2018, respectively, and are considered fully correlated between the years.

\subsection{Emittance change}
\label{sec:emitEvol}

Since the \vdM scans in $x$ and $y$ are performed sequentially, whereas Eq.~\eqref{eq:SigmaVis} assumes simultaneously measured peak rate and width values, beam parameters that shift in time can introduce a bias in the measurement of \sigmaVis.

The effect of a gradual change is evaluated using the trends of the measured bunch convolution widths, \Sigx and \Sigy, and of the head-on rates of all \vdM scans within each fill.
For each colliding bunch pair, these observables are parameterized as linear functions of time, which provide an adequate description of the data.
To abstract from the statistical fluctuations of the measured values around the trend lines, two versions of \sigmaVis are constructed using $R$, \Sigx and \Sigy parameters derived from the trend lines. 
First, $\sigma_\text{vis,1}$ is computed using parameters obtained from the fits evaluated at the times of the corresponding scan members ($t_x$, $t_y$), representing the experimental setup. Second, 
$\sigma_\text{vis,2}$ is derived by evaluating all three trends at a common reference time, defined as the midpoint of the scan pair, $t_0 =(t_x+t_y)/2$, representing the ideal scenario.
The following ratio is used to quantify the bias arising from the time dependence of the emittance:
\begin{equation}\label{eq:ec}
    \frac{\sigma_\text{vis,2}}{\sigma_\text{vis,1}} = \frac{R(t_0)\Sigx(t_0) \Sigy(t_0)}{\frac{1}{2}(R(t_x)+R(t_y))\Sigx(t_x) \Sigy(t_y)}.
\end{equation}

The cross section values obtained using the interpolation to a common reference time ($\sigma_\text{vis,2}$) are, on average, 0.15\% (0.22\%) larger in 2017 (2018) than those calculated using their respective scan times ($\sigma_\text{vis,1}$).
An exception is observed for the single 2017 scan pair in which the y scan preceded the x scan (vdM2 in Fig.~\ref{figures:orbitdrift:dorosvstime}). There $\sigma_\text{vis,2}$ is 0.23\% smaller than $\sigma_\text{vis,1}$. This reversal in the relative size is caused by the different slopes of the \Sigx and \Sigy trends.

Applying this correction improves the scan-to-scan consistency in 2017, while in 2018, where all scans were performed in the $x$-then-$y$ order, only the central value is shifted. 
The choice of reference time has a negligible impact: using $t_0 = t_x$ or $t_y$ changes the result by 0.01\% ($<$0.01\%) in 2017 (2018). 
Similarly, replacing the linear parameterization with a third-degree polynomial alters the result by less than 0.01\% in both years.
Therefore, 
no additional uncertainty is assigned.

\subsection{Orbit position}
\label{sec:OD}

In addition to the beam-beam deflection described in Section~\ref{sec:BB}, the separation of the colliding bunches can also be modified as a result of systematic shifts of the beam orbits due to nonlinear effects in the LHC magnets (``hysteresis'')~\cite{CERN-ACC-NOTE-2022-0013}, as well as by gradual orbit drift and sudden random changes of the beam positions.
To correct for a possible bias in the \sigmaVis estimation caused by such anomalous orbit movements, the orbit positions as measured with the DOROS and arc BPM systems are analyzed.

\begin{figure*}[!p]
\centering
\includegraphics[width=\textwidth]{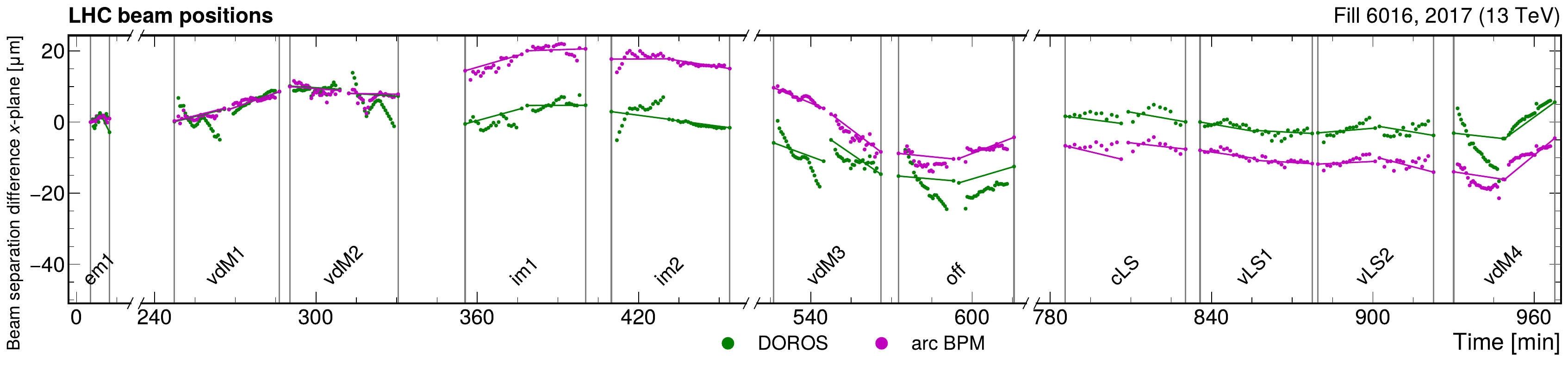} \\
\includegraphics[width=\textwidth]{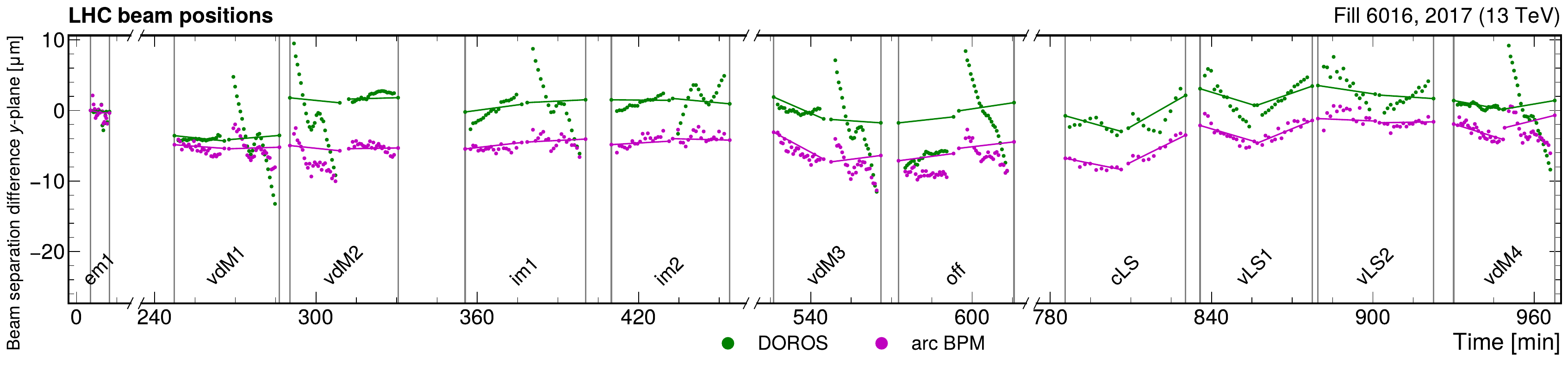} \\
\includegraphics[width=\textwidth]{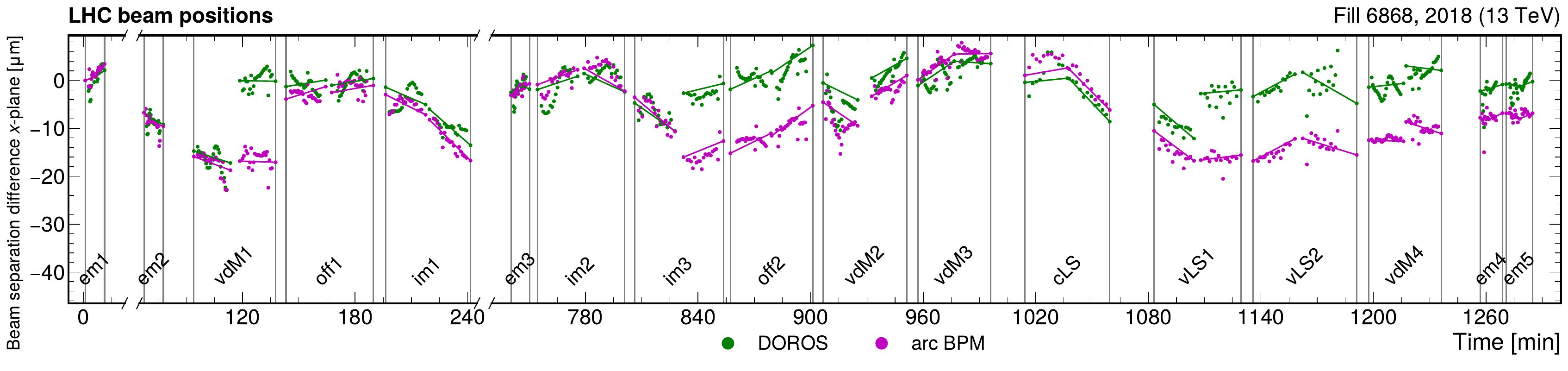} \\
\includegraphics[width=\textwidth]{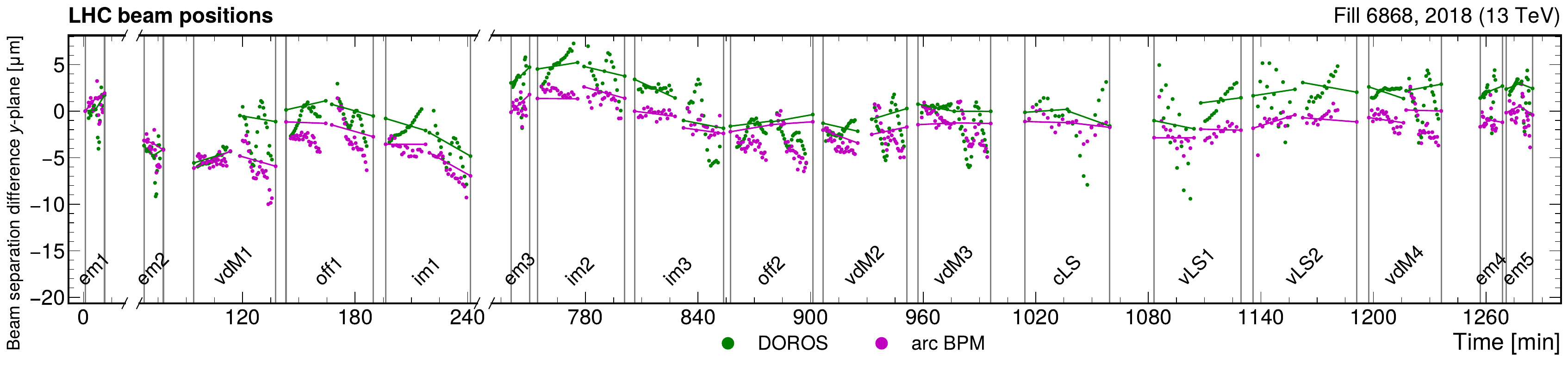}
\caption{
    Difference between nominal and measured beam separation per scan step as a function of time for the \vdM scans during LHC fills 6016 (upper two figures) and 6868 (lower two figures).
    The scan boundaries are marked by vertical lines and labeled with the 
    corresponding scan identifiers.     The solid green and magenta lines represent the linear interpolation between the head-on reference points before and after each scan, showing the linear orbit drift. The individual data points demonstrate the beam-beam deflection nicely in the scanning direction. 
    Differences between the two BPMs, DOROS and arc BPM, are expected due to differing sensitivity to the beam-beam deflection, scan-dependent offsets, and length scale differences.
}
\label{figures:linorbitdrift}
\end{figure*}

Transverse orbit positions are measured as a function of time with approximately 1\unit{s} granularity, integrated over the full orbit.
The measurements taken at each side of the IP are averaged.
These values are further averaged in each individual \vdM scan step, as well as for 30\unit{s} head-on periods just before and after each \vdM scan. The standard deviation of the corresponding per-second positions is considered as the uncertainty of the beam position for the given step in the analysis procedure that follows.
The drift is calculated with respect to the nominal positions determined by the fixed corrector magnet settings and is usually assumed to have a steady linear as well as a complex residual component during a given scan.
Figure~\ref{figures:linorbitdrift} shows the measured deviations.
As random processes also contribute, the orbit drifts are not reproducible with different scans and, if uncorrected, lead to scan-to-scan variations in \sigmaVis.

To account for gradual drift, linear orbit drift corrections are obtained by interpolating between BPM measurements taken during the head-on collision periods immediately before and after each \vdM scan.
In the scanning direction, the measured drift is applied as a correction to the beam separation, while in the nonscanning direction it is propagated to the measured luminometer rates using the bunch convolution width from the orthogonal scan.
The combined impact on \sigmaVis of the corrections in the two
directions derived from the arc BPM data is, on average, 0.13\% (0.05\%) in 2017 (2018).

The aim of residual orbit drift correction is to address all deviations not described by linear orbit drift, beam-beam deflection, or LS differences between the BPMs and the nominal positions. These LS differences originate from inconsistencies between the calibration of the BPM system and that of the steering magnets at the precision relevant for the analysis. This effect is detailed in Section~\ref{sec:LS}.
The residual orbit drifts in the scanning and nonscanning directions, \DelU{residual} and \DelV{residual}, are extracted from the BPM measurements in the scanning and nonscanning directions, \posU{BPM} and \posV{BPM}, as the residuals of the following fit model:
\begin{equation}\begin{aligned}\label{eq:od}
    \DelU{residual} &= \posU{BPM} - \posU{linOD} - \alpha\posU{nom} - \delta\posU{BB}, \\
    \DelV{residual} &= \posV{BPM} - \posV{linOD} - \gamma\posU{nom} - \frac{\gamma}{\alpha}\delta\posU{BB}.
\end{aligned}\end{equation}
In both equations, the second term on the right-hand side accounts for the linear orbit drifts, in the scanning (\posU{linOD}) and nonscanning (\posV{linOD}) directions, measured as described above.
The third term in the first equation accounts for the nominal movements (\posU{nom}) and any potential LS differences between the BPM measurements and the nominal positions, parametrized by a LS factor $\alpha$.  The third term in the second equation stands for a potential geometric misalignment between the BPM and nominal displacement coordinate systems. This could lead to a ``leakage'' of changes from the scanning nominal direction into the nonscanning BPM direction (parametrized by a leakage factor $\gamma$).
The fourth term accounts for the beam-beam deflection (\posU{BB}) as calculated in Section~\ref{sec:BB}, scaled in the scanning direction with a factor $\delta$ to account for the fact that only colliding bunch pairs undergo deflection, whereas the BPM measurement is averaged over all bunches, including noncolliding ones.
In the nonscanning direction, beam-beam deflection enters only through leakage.
The fraction of colliding bunches to all bunches is about 62\% in 2017 and 89\% in 2018.

\begin{figure}[!htpb]
\centering
\ifthenelse{\boolean{cms@external}}{
\includegraphics[width=0.45\textwidth]{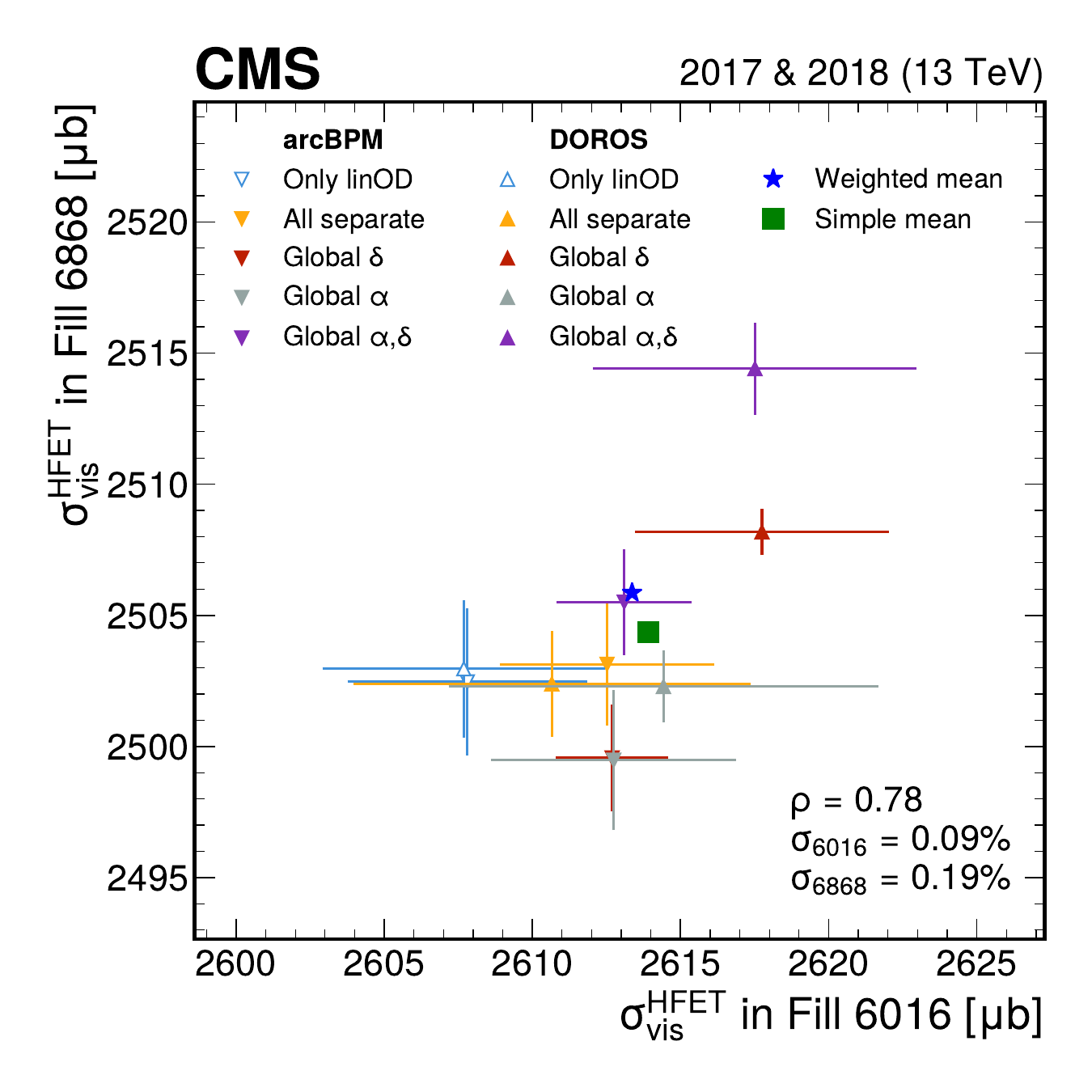}
}{
\includegraphics[width=0.65\textwidth]{Figure_005.pdf}
}
\caption{
    Visible cross section estimates for the HFET luminometer with all corrections applied for the 2017 and 2018 calibration fills (6016 and 6868). 
    Results are shown for various orbit drift correction methods, corresponding to different assumptions of the $\alpha$ and $\delta$ parameters in the residual orbit drift fits. These parameters describe the BPM (arc or DOROS) length scale and the dilution of the beam--beam deflection, respectively. 
    The markers represent the average over all scans, while the error bars correspond to the standard deviation, reflecting scan-to-scan reproducibility. 
    The baseline method (purple downward-facing triangle) uses arc BPM measurements and assumes common $\alpha$ and $\delta$ values for all scans within a given year. 
    The corresponding \sigmaVis value is closest to the mean of all residual orbit drift configurations (blue star and green square) and is taken as the central correction method.
    The empty triangles, representing the linear-only orbit drift corrections, are excluded from the averages and from the other metrics shown.
    The correlation between the years, shown in the lower right corner, is large (0.78). The standard deviations, taken as the uncertainty associated with the choice of method, are also indicated. The difference in \sigmaVis between the two years is consistent with the expected efficiency loss of HFET due to radiation-induced aging. 
   }
\label{figures:ODcorrelation}
\end{figure}

The scanning and nonscanning directions are fitted simultaneously.
The model can be fitted either separately for each scan, or in a global fit with one or more of the parameters set to be the same for all scans.
Although the BPM length scales and the beam-beam deflection scale factors are expected to be stable throughout a fill, the fitted values of $\alpha$ and $\delta$ demonstrate variations when fitted for each scan individually.
The $\gamma$ parameter is considered independent for each scan.
In Fig.~\ref{figures:ODcorrelation}, the \sigmaVis results for 2017 and 2018 are compared with different orbit drift corrections applied.
Results obtained with only linear orbit drift corrections applied are shown for reference, while results including both linear and residual orbit drift corrections are compared for four different fit configurations.
In all cases, separate sets of results are presented using the DOROS and arc BPM data.
The large correlation of 0.78 between the results for the two years indicates that the differences stem from systematic effects and not from statistical fluctuations.

\begin{figure*}[!htpb]
\centering
\includegraphics[width=0.95\textwidth]{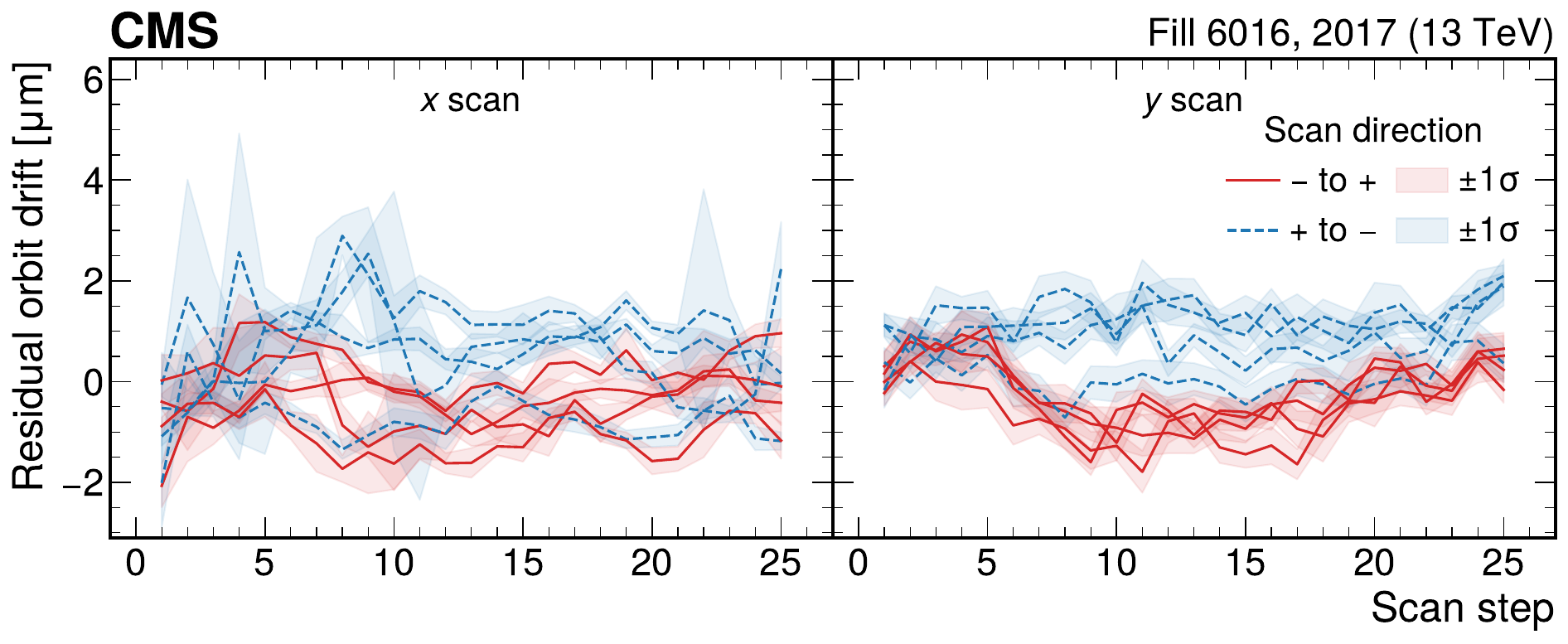} \\
\includegraphics[width=0.95\textwidth]{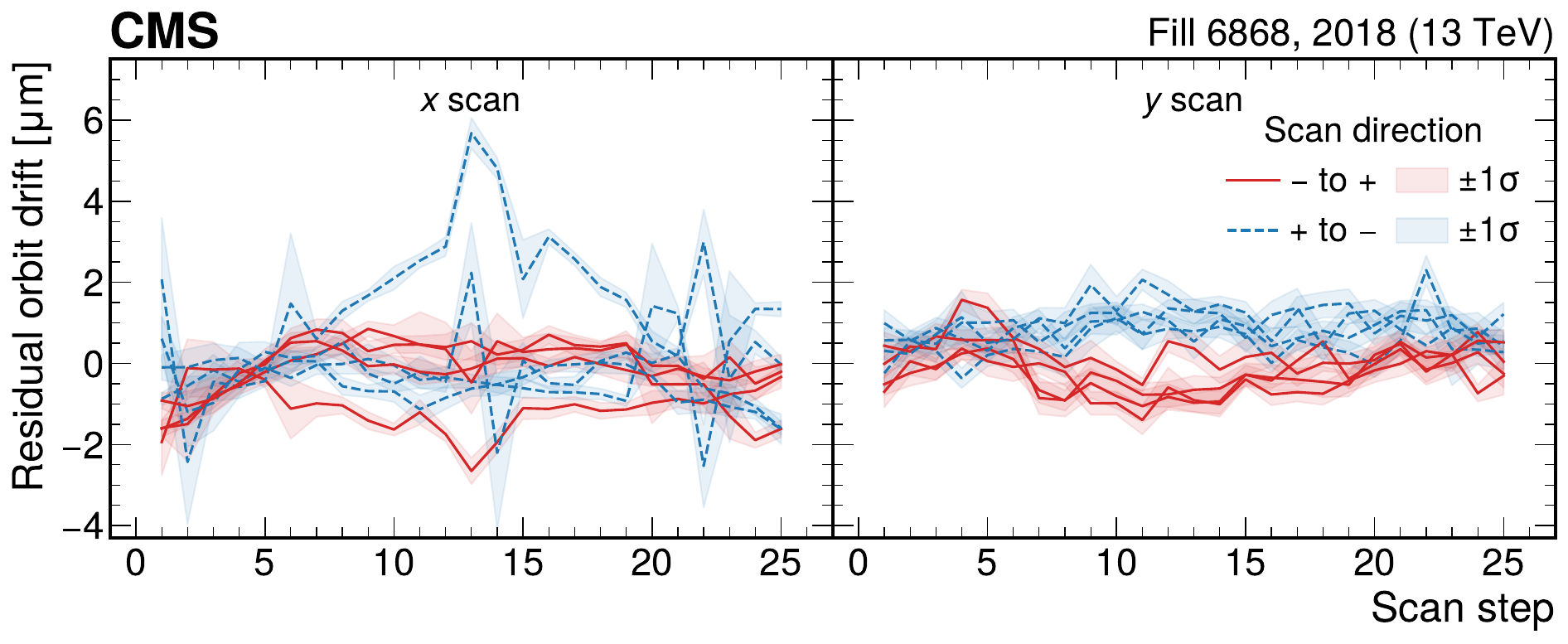}
\caption{
    Residual orbit drift corrections to the single-beam positions as a function of the scan step number for each regular \vdM scan, as derived from the arc BPM measurements using a fit with common $\alpha$ and $\delta$ parameters for all scans within a year, shown for 2017 (upper) and 2018 (lower).
    The red (blue) lines correspond to residuals for the beam moving from negative (positive) to positive (negative) direction in the transverse plane with respect to the LHC reference frame, as illustrated in Fig.~\ref{figures:orbitdrift:dorosvstime}.
    This convention is adopted to enhance the visibility of reproducible features, e.g., those arising from magnetic nonlinearities.
    The uncertainty bands represent the standard deviation of the individual BPM readings within each scan step.
}
\label{figures:resorbitdrift}
\end{figure*}

Among the considered configurations, the results obtained using orbit drift corrections derived from arc BPM data, with both $\alpha$ and $\delta$ treated as global parameters, are closest to the mean of all the results, and exhibit a good scan-to-scan agreement for both years (small error bars).
The residual orbit drift in the scanning direction for this fit scenario is shown in Fig.~\ref{figures:resorbitdrift}.
This set of corrections is used for the \sigmaVis determination, with an average impact (range of impacts) of 0.20 (between $-0.90$ and 0.53)\% in 2017 and 0.13 (between $-0.54$ and 0.36)\% in 2018.
The standard deviation of all fit scenarios is used to assign a systematic uncertainty of 0.09\% (0.19\%) in 2017 (2018), which also covers the uncertainty in the linear orbit drift correction.
This uncertainty is treated as fully correlated between the two years.

\subsection{Length scale}
\label{sec:LS}

In the vdM scans, the beams are steered over a significantly wider range of transverse displacements than in normal LHC operations.
While the LHC magnets are calibrated for the whole range of beam displacements, the precision can be improved with in-situ measurements using the CMS tracker~\cite{TRK-11-001} coordinate system as the distance reference.
The goal of the length scale calibration is to use vertices reconstructed with the CMS tracker to determine a correction factor, \lsTrNom, 
relating the nominal beam separation, as inferred from the magnet current settings, to the separation in the tracker frame.
This LS factor between tracker and nominal positions is different from the LS factor $\alpha$ between BPM and nominal positions discussed in Section~\ref{sec:OD}, which we will refer to as \lsBpmNom in this section.
In all cases, separate LS factors are derived for the $x$ and $y$ directions.

The LS calibration is performed with two different sets of beam-separation scans, referred to as constant- (cLS) and variable-step size (vLS) length scale scans.

In the cLS method, which was also analyzed in Ref.~\cite{CMS:2021xjt}, the two beams are held at a constant separation corresponding to the expected bunch convolution width, and moved in several steps along the $x$ or $y$ direction.
A linear fit to the average transverse position of reconstructed vertices as a function of the average nominal position of the two beams yields the average LS of the two beams as the slope.
The cLS scans are performed as $x$-$y$ pairs, in both the ``forward'' and ``backward'' directions. A separate analysis of the forward and backward scans provides an internal cross-check for reproducibility and direction dependence, which may arise from magnetic nonlinearities.

In the vLS method, performed in CMS for the first time in 2017, one beam is moved in steps along the $x$ or $y$ direction. At each position, a three-step mini-scan is performed with the other beam along the same axis.
A Gaussian fit to the rates, together with a linear fit to the mean vertex positions as a function of the beam separation during each mini-scan, is used to determine the head-on position in the tracker frame. This position is obtained by evaluating the linear fit at the Gaussian maximum. As this also corresponds to the tracker coordinates of the scanned beam, the LS factor can be extracted from a linear fit between the nominal and tracker coordinates of the scanned beam. For each transverse direction, the separation LS is taken as the average of the LS factors of the two beams. The vLS scans are performed in sets of four, one for each beam in both $x$ and $y$.

For both the cLS and the vLS, orbit drift corrections derived from either the DOROS or the arc BPM measurements are applied. The correction procedure consists of computing the average measured step size, optionally after an outlier removal, and adjusting each of the nominal step sizes by the difference between the BPM-measured and the average step size~\cite{CMS:2021xjt}. This procedure cancels possible BPM LS effects to first order.
Since the cLS and vLS methods are affected differently by the various sources of uncertainties, they are complementary.

To reduce the uncertainty stemming from the orbit drift during the LS scans, we present for the first time an alternative approach in the analysis of the two LS calibration methods.
The improved approach proceeds in two steps. First, the BPM LS factors \lsBpmNom are measured as in Section~\ref{sec:OD}, relying on the full \vdM scan program and benefiting from the averaging over many scan points, which effectively reduces the impact of orbit drifts.
Second, the LS scans are used to measure length scale factors \lsTrBpm between the BPM measurements and the tracker, where orbit drift effects largely cancel as both instruments see the same beam movements.
The final result is then the product of the two LS factors, $\lsTrNom=\lsTrBpm\,\lsBpmNom$.

\begin{figure*}[!htbp]
\centering
\includegraphics[width=0.45\textwidth]{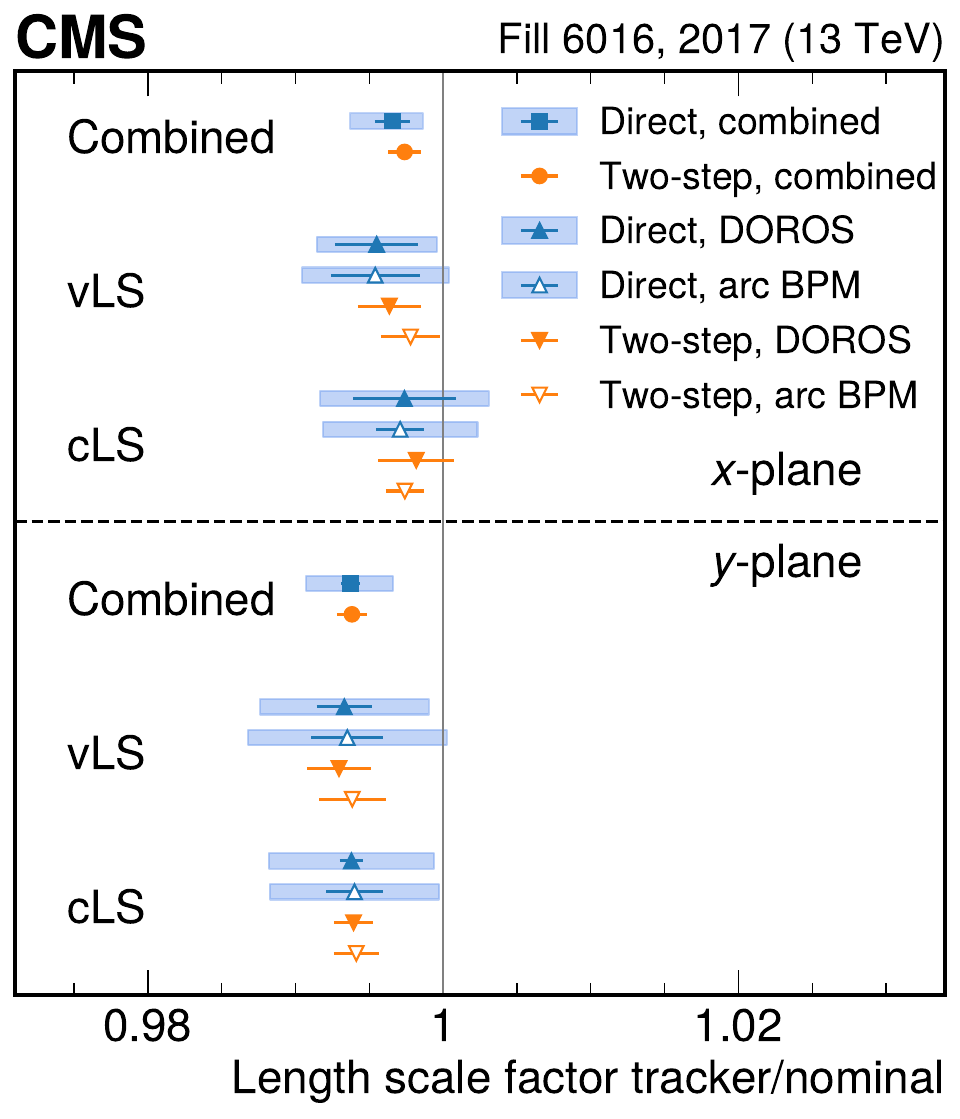}
\hspace{0.05\textwidth}
\includegraphics[width=0.45\textwidth]{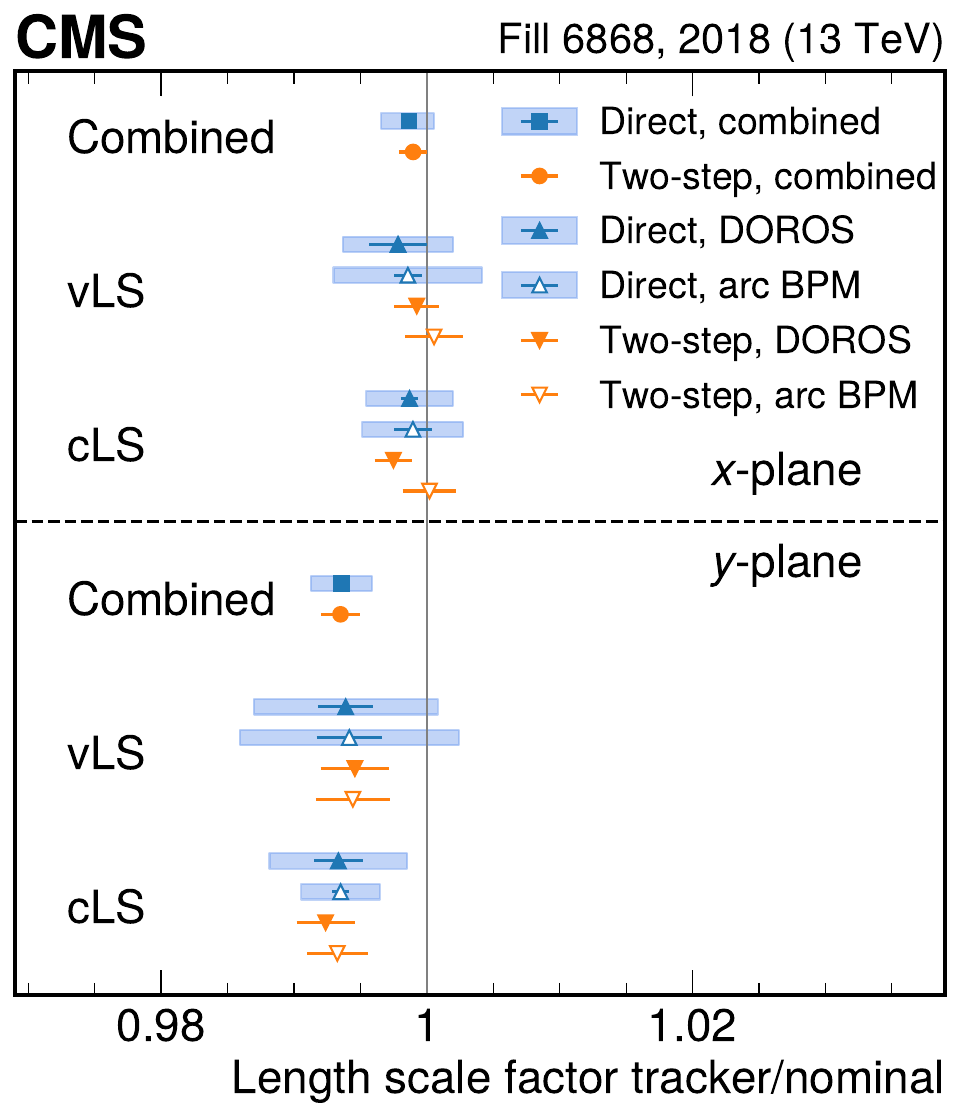}
\caption{%
    Length scale factors \lsTrNom obtained with the direct (blue squares and upward triangles) and two-step (orange points and downward triangles) approaches for 2017 (left) and 2018 (right). Results are shown separately for cLS and vLS scans, using either the DOROS (filled triangles) or arc (empty triangles) BPMs, as well as their combination (squares and circles).
    For the direct approach, the error bars represent the total uncertainty, with the orbit drift component estimated from the difference between results obtained with and without outlier removal, likely underestimating this source. The shaded bands also denote the total uncertainty, where the OD contribution is instead computed as the standard deviation of the measured step sizes, providing a conservative estimate.
    Since the combination of the vLS and cLS results is computed as a weighted mean, the shaded band is not necessarily centered on the blue squares.
    For the two-step approach, the error bars represent the total uncertainty.
}
\label{fig:lengthscale}
\end{figure*}

The results of the direct and the two-step LS measurements are shown in Fig.~\ref{fig:lengthscale}, comparing outcomes obtained using the constant and variable LS scans and the DOROS- and arc BPM-based measurements.
For the direct estimation of \lsTrNom, the error bars indicate the total uncertainty, obtained from adding in quadrature the statistical uncertainty from the linear fits (typically $<$0.04\% for cLS, 0.10\% for vLS), the effect of tracker alignment (typically $<$0.04\%, 0.12\%), the difference between results with and without outlier removal in the orbit drift correction (typically 0.12\% for cLS, 0.2\% for vLS), and---in the cLS case---the difference between the results from the forward and backward movements (typically $<$0.1\%).
The shaded bands represent the same estimate of the total uncertainty, but with a modified treatment of the orbit drift (OD) component. Instead of using the difference between results obtained with and without outlier removal, the OD uncertainty is taken as the relative standard deviation of the step sizes measured by the BPMs during the corresponding scans. This quantity has been identified in simulation studies as a good estimate of the OD uncertainty for random walk scenarios
in the direct approach.
This component amounts to 0.49 and 0.34\% for the cLS measurements and 0.66 and 0.62\% for the vLS measurements in 2017 and 2018, respectively.

For the two-step estimation of \lsTrNom, the error bars indicate the uncertainties in \lsTrBpm and \lsBpmNom added in quadrature.
The relevant uncertainties in \lsTrBpm are the statistical uncertainty of the linear fits (typically 0.03\% for cLS, 0.1\% for vLS), the effect of tracker alignment (typically 0.05\% for cLS, 0.1\% for vLS), and---in the cLS case---the difference between the results from the forward and backward movements (typically $<$0.05\%).
For \lsBpmNom, we employ the weighted average and standard deviation of the LS factors obtained in individual per-scan fits as the central value and uncertainty, respectively.

The size of potential instrumental effects in the BPM measurements can be appreciated by comparing the results based on DOROS and arc BPM data: the agreement is good.
The comparison of the cLS and vLS results also provides information on the impact of random orbit drift, since they rely on independent scans.
For each approach, the results from both scan types and using either BPM are combined through a weighted average.
Overall, the central values from the two approaches agree well.
The two-step approach is expected to be more robust against orbit drift biases, and for this reason it is used for the \vdM scan analysis.
The individual LS factors are $0.9938\pm0.0010$ and $0.9974\pm0.0010$ in 2017 and $0.9933\pm0.0015$ and $0.9990\pm0.0011$ in 2018, respectively, in $x$ and $y$.
The resulting correction on the visible cross section is $(-0.88\pm0.15)\%$ in 2017 and $(-0.75\pm0.18)\%$ in 2018.

\subsection{Transverse factorization of bunch proton densities}
\label{sec:factorization}

The \vdM method assumes that the bunch proton densities are factorizable in the transverse directions (Eq.~\eqref{eq:fact}).
When this assumption does not hold, the estimation of \sigmaVis via the product of the bunch convolution widths measured in two orthogonal directions, described in Eq.~\eqref{eq:SigmaVis}, is potentially biased.
Despite a dedicated tailoring of the proton bunches in the injection chain aimed at minimizing non-Gaussian bunch proton densities at the LHC~\cite{CERN-ACC-NOTE-2013-0008}, biases at the percent level are still possible, leading to one of the dominant uncertainty sources in the luminosity calibration.
In this work, two approaches are employed to measure and correct for this factorization bias. The first is the luminous region (LR) analysis, which reconstructs the bunch proton densities directly and was first used by CMS in Ref.~\cite{CMS:2021xjt}. The second is the  2D rate fit (2D-RF) method using off-axis (here: offset) scans, which fits the convolved bunch shape and was originally developed for Ref.~\cite{LUM-20-002}. The two methods are used to validate each other.

The LR analysis, which provides the central value of the correction, is an adaptation of the method described in Ref.~\cite{Webb:2020875} with various improvements~\cite{ThesisPMajor2024} to increase the robustness of the fit and introduce corrections for the beam-beam, orbit drift, and emittance change effects.
The method relies on vertex data that are collected using zero-bias triggers gated on only five of the colliding bunch pairs.
This restriction is necessary for accumulating enough data per scan step for each studied bunch pair within the available trigger bandwidth.
Unlike the 2D rate fit method, the LR analysis does not require special scans; 
it can be used with 
standalone vdM or beam-imaging scan pairs, 
or with a combination of consecutive scan pairs which could include offset scans as well. As a result, the LR method provides corrections at the level of individual \vdM scans and enables a determination of the time dependence of nonfactorization with high granularity.

The spatial distribution of primary vertices, referred to as the luminous region, is independently fitted at each scan step with a 3D Gaussian function convolved with a vertex resolution kernel~\cite{TRK-11-001}. This procedure provides the nine LR parameters: three describing the position of the center, three the widths, and three the orientation.

To describe the 3D proton density distributions of the bunches, the LR method employs fit models based on the sum of up to five 3D Gaussian functions with nondiagonal covariance matrices.
The model parameters are determined through a simultaneous fit to the luminometer rate and to the nine LR parameters as a functions of the beam positions using all scan steps in the data set considered.

The length ratio of the two bunches is constrained based on the LDM bunch length measurements.
Possible inaccuracies of the vertex resolution kernel are modeled using a floating term in a quadratic sum with the computed transverse LR widths.
The used orbit position data are corrected using the LHC BPMs, taking beam-beam deflection and the per-beam LS into account. The effects of beam-beam optical distortions and gradual emittance evolution are incorporated by modulating the bunch model widths on a per-step basis.

\begin{figure*}[!htpb]
\centering
\includegraphics[width=0.48\textwidth]{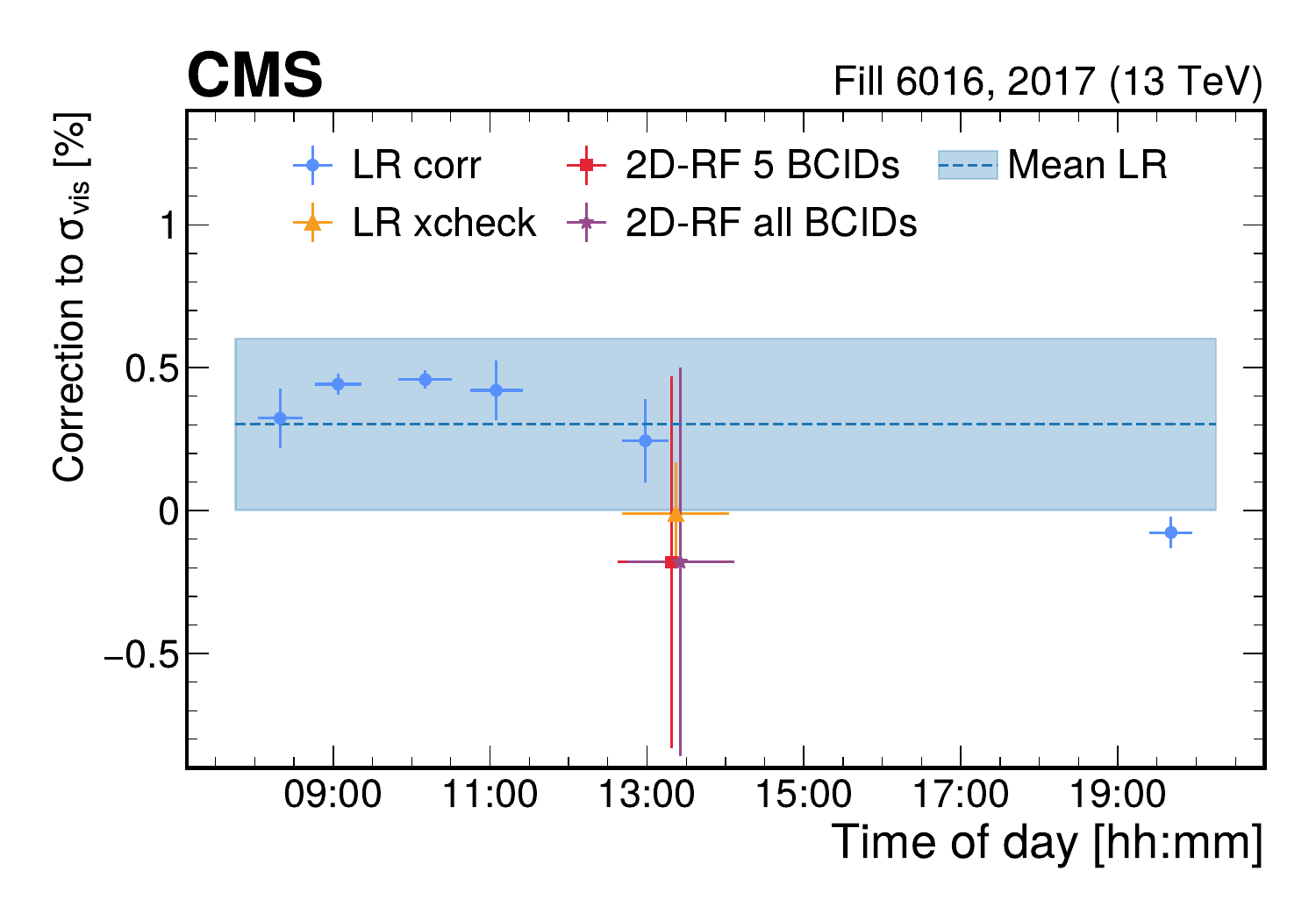}
\includegraphics[width=0.48\textwidth]{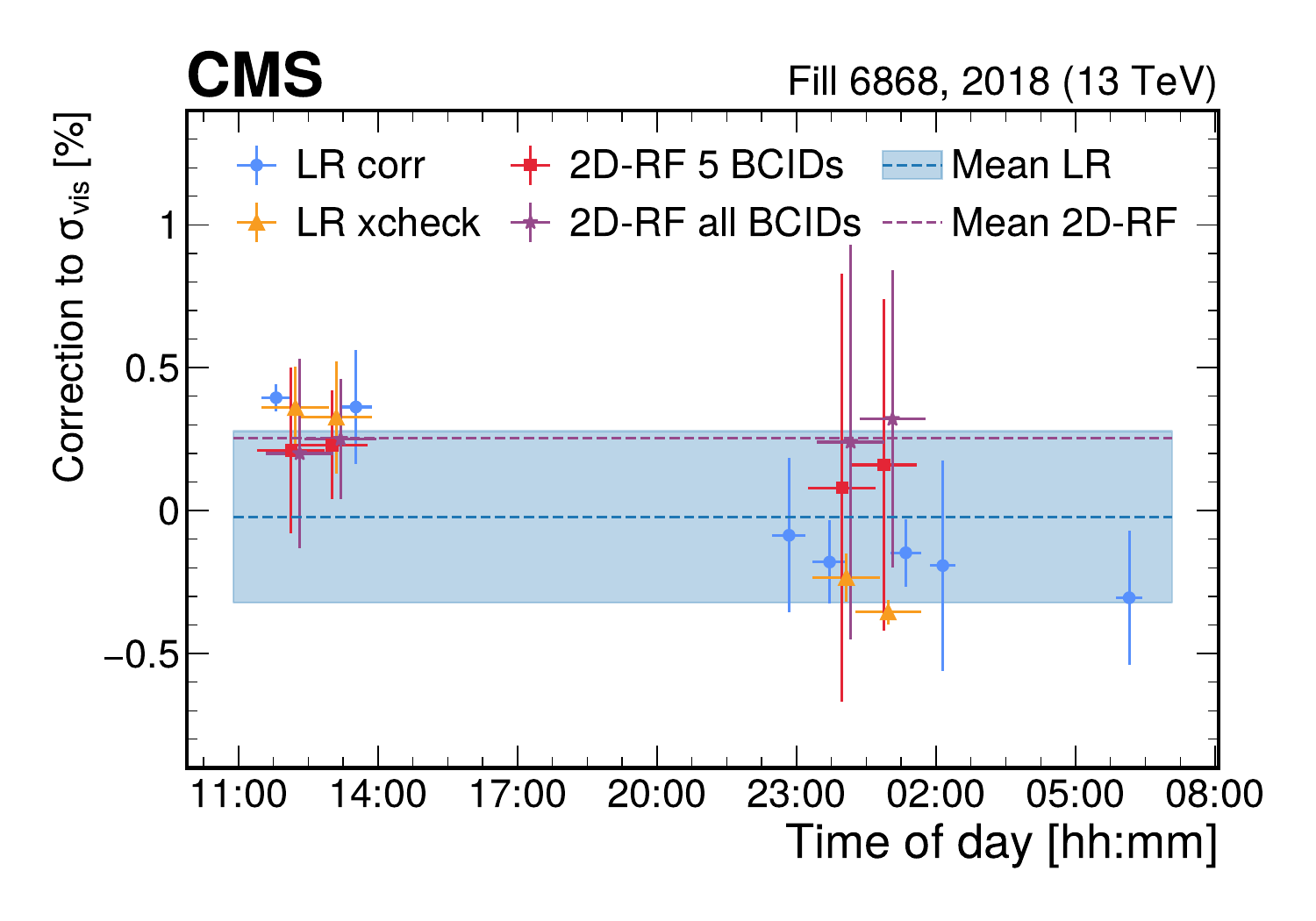}
\caption{
    Nonfactorization estimates for 2017 (left) and 2018 (right) using the luminous region (LR) and 2D rate fit (2D-RF) methods.
    For all markers, the horizontal error bars indicate the time period during which the input data for the fit were collected.
    For the LR method (blue), both per-scan and average corrections are shown. The vertical error bars correspond to the standard deviation of the BCIDs, while the markers show the average value. 
    The shaded band represents the total uncertainty of the LR method, dominated by the closure uncertainty. The final correction is based on the per-scan LR measurement.
    The orange markers indicate LR results obtained using the same input scans as the 2D-RF method, i.e., a combination of one on-axis and one off-axis scan pair.
    The red and green markers show the corrections derived using the 2D-RF method for the five BCIDs used in the LR fit and for all colliding BCIDs, respectively. The vertical error bars on these markers reflect the full uncertainty associated with the 2D-RF method. The difference between the two methods is included in the total factorization uncertainty. 
}
\label{fig:lumr:corr}
\end{figure*}

The correction and its uncertainty for the factorization assumption are calculated by simulating vdM experiments based on the best fit model for the bunch particle densities.
In each experiment, the vdM results are compared to the ``true'' beam convolution integral of the model to study the bias.
This procedure also accounts for potential systematic effects caused by the choice of the vdM fit function.
The corrections to the visible cross section are shown as functions of time in Fig.~\ref{fig:lumr:corr}.
The values are applied scan-by-scan based on the corresponding single scan pair fit results.
The mean impact of the correction is 0.36\% (0.04\%) in 2017 (2018). 
This value differs slightly from the simple average of the corrections shown in Fig.~\ref{fig:lumr:corr}, as the uncorrected \sigmaVis values act as weights for the impact.

The systematic uncertainty of the derived corrections is dominated by the possible intrinsic bias of the method (0.25\%).  This potential bias is studied using Monte Carlo simulation in a closure test in which the ``true'' and ``measured'' nonfactorization corrections are compared for a wide range of bunch proton densities.
Recent improvements to the method enhanced its precision by a factor of two compared to Ref.~\cite{CMS:2021xjt}.
This important source of uncertainty is not considered in Ref.~\cite{ATLAS:2022hro}.
Several additional sources of uncertainty related to the fit model, the various inputs, and corrections that affect them were investigated.
The only significant effect arises from the parametrization of the beam spot ellipsoid, and whether the orientation parameters of the bunch density components are shared or are independent.
These two sources are anti-correlated and contribute 0.10\% in total. 

To check the modeling of the tails, the LR fits are carried out incorporating off-axis scans into the LR fit, when available. The resulting differences between the on-axis only and the combined fits, averaged over the two years, yield 0.10\%, which is assigned as an additional uncertainty.
Combining these sources amounts to a final LR uncertainty of 0.29\%.

The 2D rate fit method~\cite{ThesisPMajor2024} uses combinations of on-axis (vdM or im) and off-axis (offset, named ``off'' in Fig.~\ref{figures:orbitdrift:dorosvstime}) scan pairs, that were conducted close in time, to directly determine the bunch convolution shape.
Offset scans are vdM-like scans with a fixed nonzero separation between the two beams in the nonscanning direction, hence sampling the tails of the rate function ($R(\Delx,\Dely)$).
The normalized rates from the luminometers measured during the on- and off-axis scan pairs are fit simultaneously as a function of the transverse beam separation ($\Delx$, $\Dely$) with 2D analytic functions that include linear correlations and/or non-Gaussian features.
The bias is determined from simulated vdM scans, with the fitted surfaces taken as the true $R(\Delx,\Dely)$.
Here, the integral of the 2D fit function is compared to the result from the simulated vdM fit.

The dominant uncertainty for the 2D-RF analysis originates from the lack of a priori knowledge of the bunch convolution shape.
Various 2D functions are considered: single and double Gaussian, the latter in various alternative parametrizations that restrict the number of free parameters, super-Gaussian, q-Gaussian, and poly$N$-Gaussian where the radial component of the Gaussian is multiplied by a symmetric $N^\text{th}$ degree polynomial.
The complexity of the functions (\ie, the number of free parameters) is restricted by the number and configuration of the measurement points.
The stability of the fits is tested by varying the relative alignment of the scan pairs and by changing the luminometer used.
The closure of the method is primarily driven by the statistical uncertainty of the peak rate and is better than 0.1\% for identical truth and fit shapes. The results are in good agreement with the LR method and are shown in Fig.~\ref{fig:lumr:corr}. 
The 2D-RF method does not rely on vertex information and can therefore use data from any colliding bunch pair.
The bias arising from the BCID selection in the LR method can be estimated by comparing the 2D-RF correction obtained for the five BCIDs used in the LR fit with that obtained for all BCIDs.
The derived corrections agree to 0.01\% for the 2017 and early 2018 scans, and to 0.16\% for the late 2018 scans. For 2018, an uncertainty of 0.16\% is assigned to the BCID selection. 
Adding this in quadrature to the LR uncertainty, the final uncertainty on the nonfactorization measurement is 0.29\% (0.33\%) for 2017 (2018), treated as fully correlated between the years.

\subsection{Visible cross section results}
\label{sec:sigmaVis}

Applying all corrections as described above, \sigmaVis is derived independently for HFET, HFOC, PCC, and PLT, for each colliding bunch pair and each of the four vdM and two (2017) or three (2018) imaging scan pairs.
Table~\ref{tab:vdM_correction} summarizes the effect of these corrections for HFET. Apart from the background contribution, the corresponding corrections have the same value for the remaining luminometers at the sub-per mille level.

\begin{table}[!ht]
\centering
\topcaption{
    The impact of the corrections applied in the vdM calibration procedure on the final (\ie, bunch- and scan-averaged) \sigmaVis value for HFET in 2017 and 2018. }
\label{tab:vdM_correction}
\renewcommand{\arraystretch}{1.1}
\begin{scotch}{lcc}
    Correction & 2017 [\%] & 2018  [\%]\\
    \hline
    Background & \NEG0.22 & \NEG0.17 \\
    Orbit drift---scanning direction & 0.05 & 0.00 \\
    Orbit drift---nonscanning direction & 0.08 & 0.05 \\
    Beam-beam deflection & 1.72 & 1.61 \\
    Dynamic-$\beta$ & \NEG1.09 & \NEG1.00 \\
    Length scale & \NEG0.88 & \NEG0.75 \\
    Residual orbit drift & 0.20 & 0.13 \\
    Nonfactorization & 0.36 & 0.04 \\
    Emittance change & 0.15 & 0.22 \\
    \hline
    Total & 0.37 & 0.13 \\
\end{scotch}
\end{table}

\begin{figure*}[!ht]
\centering
\includegraphics[width=0.45\textwidth]{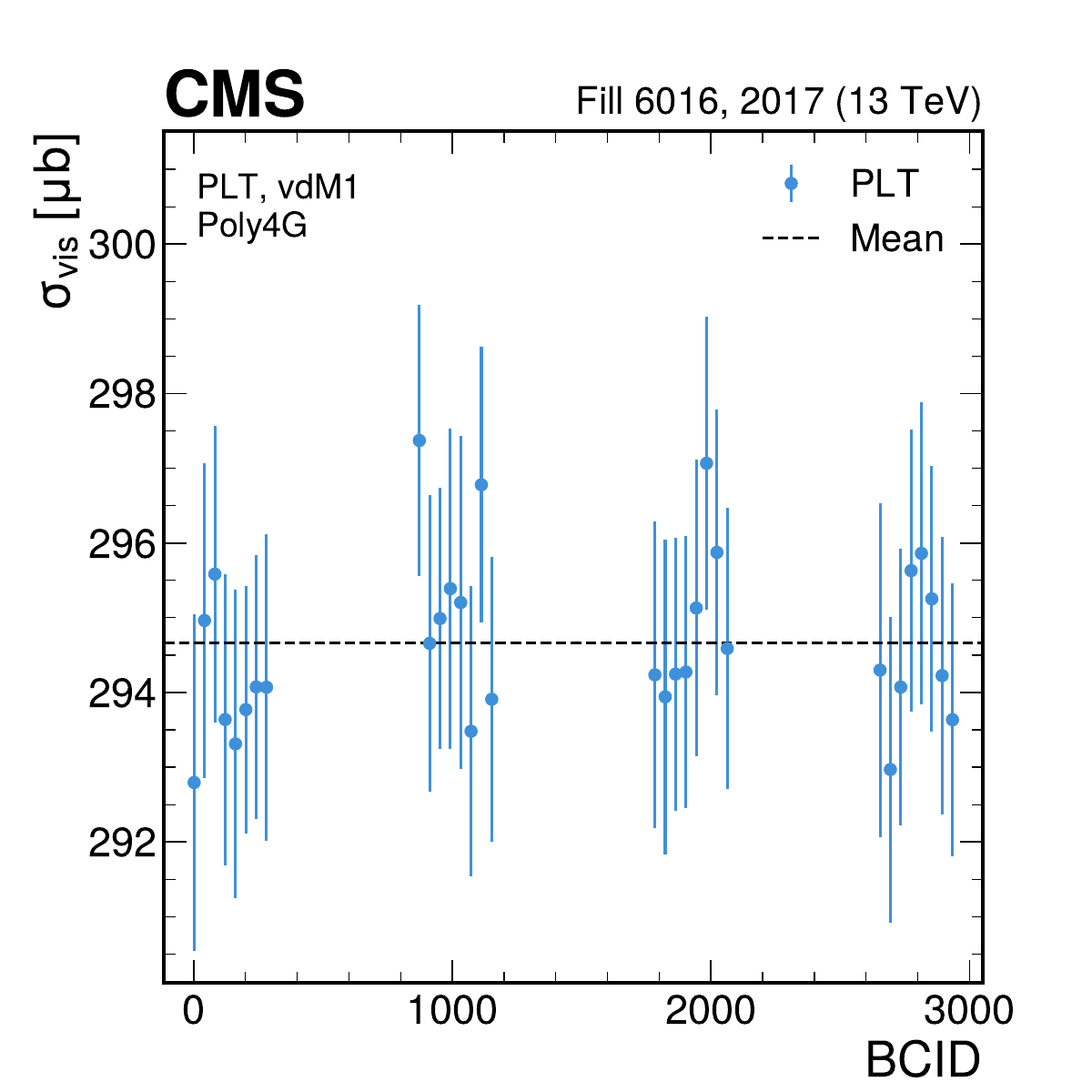}
\hspace{0.05\textwidth}
\includegraphics[width=0.45\textwidth]{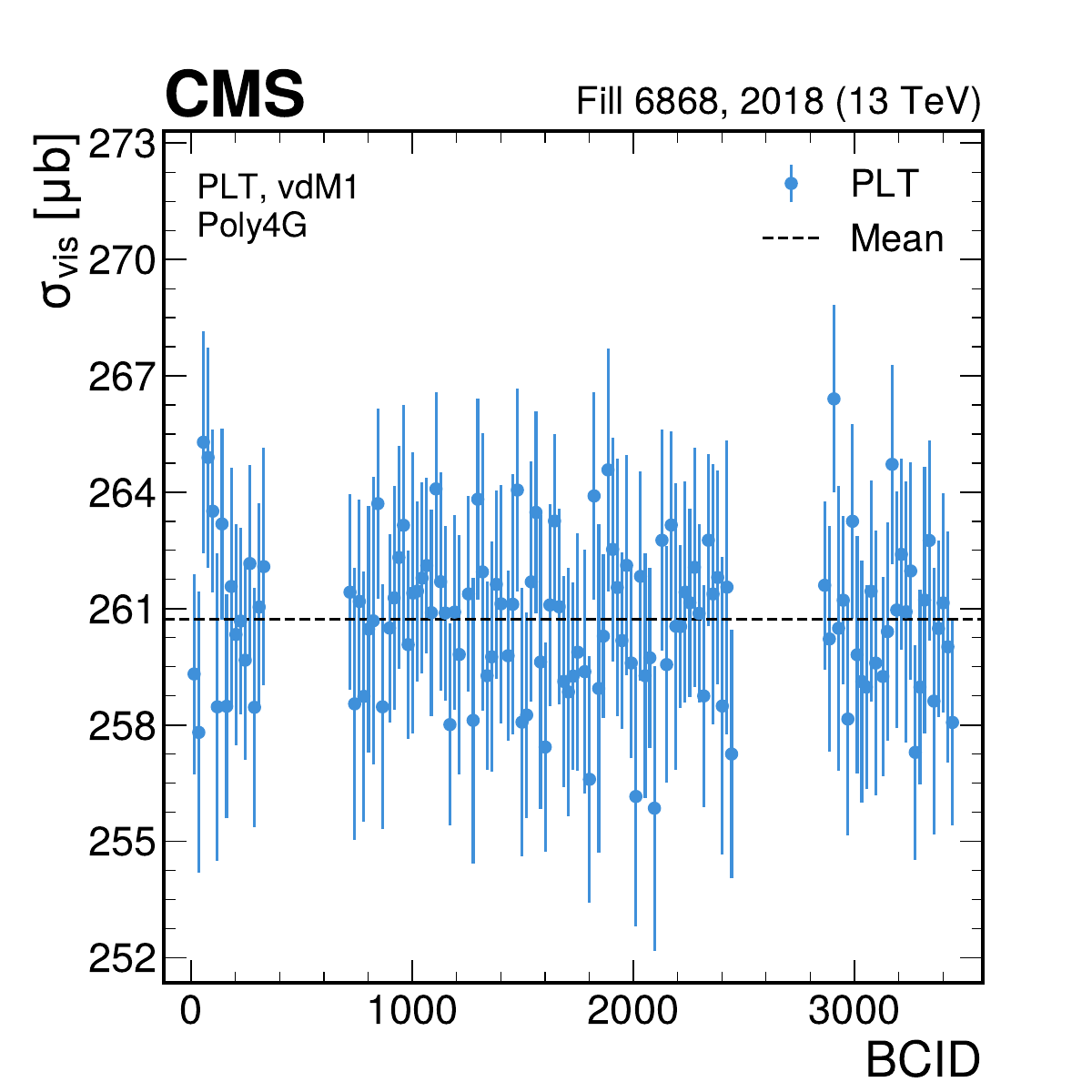}
\caption{
    Per BCID \sigmaVis values in the first scan in 2017 (left) and 2018 (right) as measured for the PLT luminometer.
    The error bars represent the uncertainty propagated from the fit.
}
\label{fig:vdM_fit_sigVis_bcid}
\end{figure*}

Figure~\ref{fig:vdM_fit_sigVis_bcid} shows the per BCID \sigmaVis values in the first scan in 2017 (left) and 2018 (right) for PLT.
The relative error on the mean falls within 0.1\% for all luminometers; this value is assigned for the uncertainty associated to the bunch-to-bunch variation of the measured visible cross sections.

\begin{figure*}[!htpb]
\centering
\includegraphics[width=0.45\textwidth]{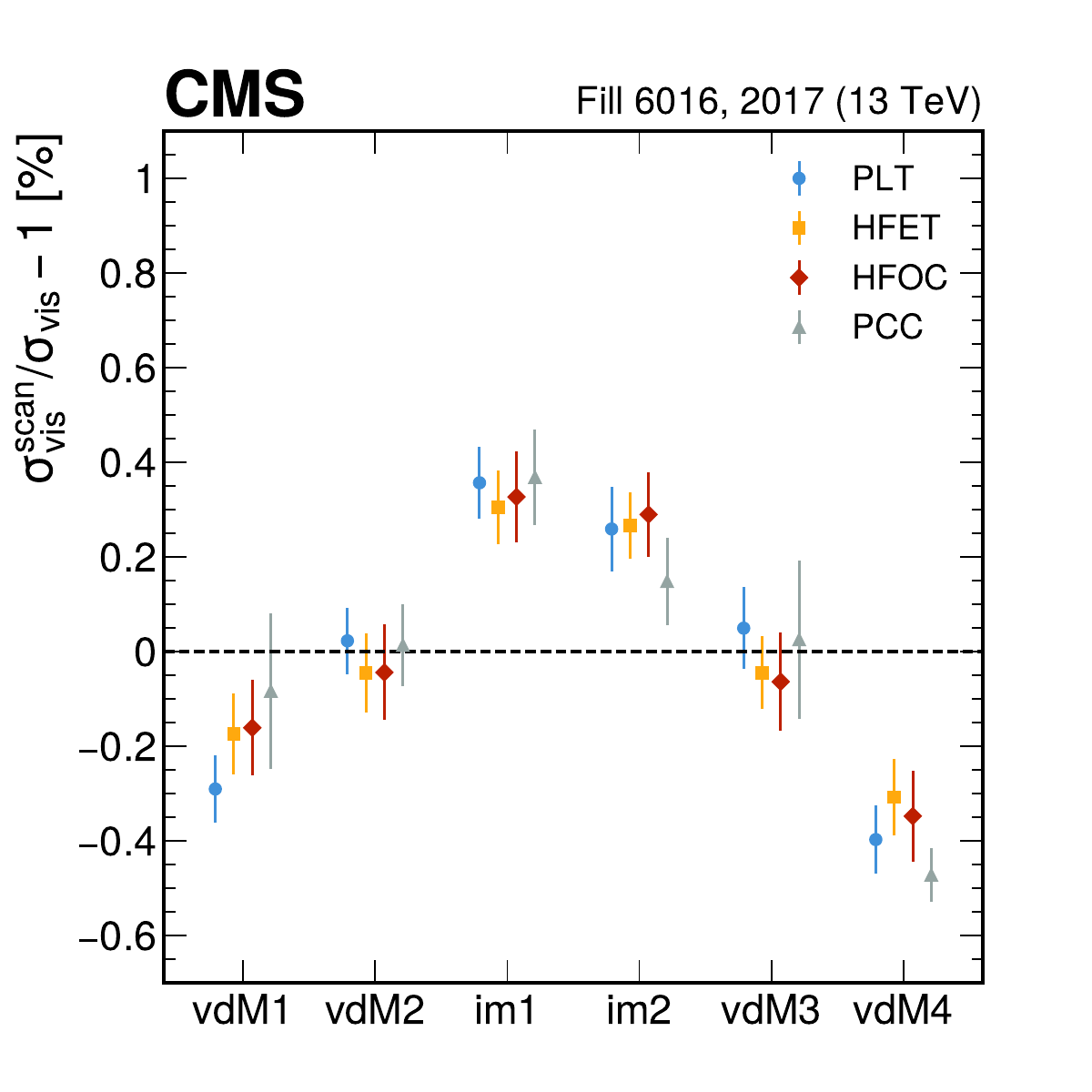}
\hspace{0.05\textwidth}
\includegraphics[width=0.45\textwidth]{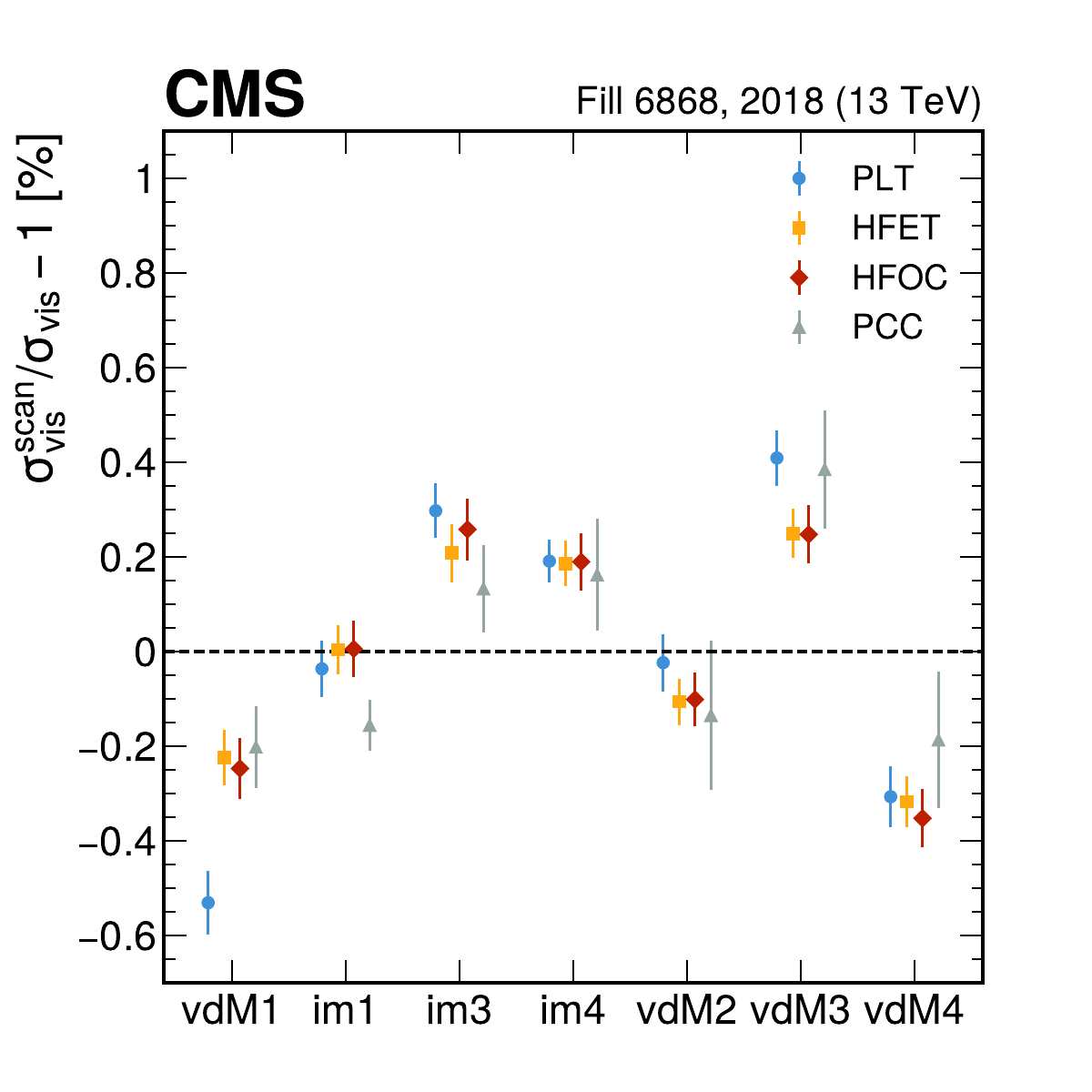}
\caption{
    The relative deviation of the BCID-averaged \sigmaVis values divided by the average value taken over all scans for all independently calibrated detectors and all scans using the Poly4G fit function in 2017 (left) and 2018 (right).
    The error bars signify the standard deviation over the BCIDs divided by the square root of the number of BCIDs.
}
\label{fig:vdM_fit_sigVis_All}
\end{figure*}

Figure~\ref{fig:vdM_fit_sigVis_All} illustrates scan-by-scan variation of the measured \sigmaVis after all corrections.
The \sigmaVis values are normalized by their average value taken over all scans to bring them to the same scale.
The standard deviation of the scans averaged over the four detectors (0.26\% in 2017 and 0.27\% in 2018) is taken to be the scan-to-scan reproducibility uncertainty.

\begin{figure*}[!htpb]
\centering
\includegraphics[width=0.475\textwidth]{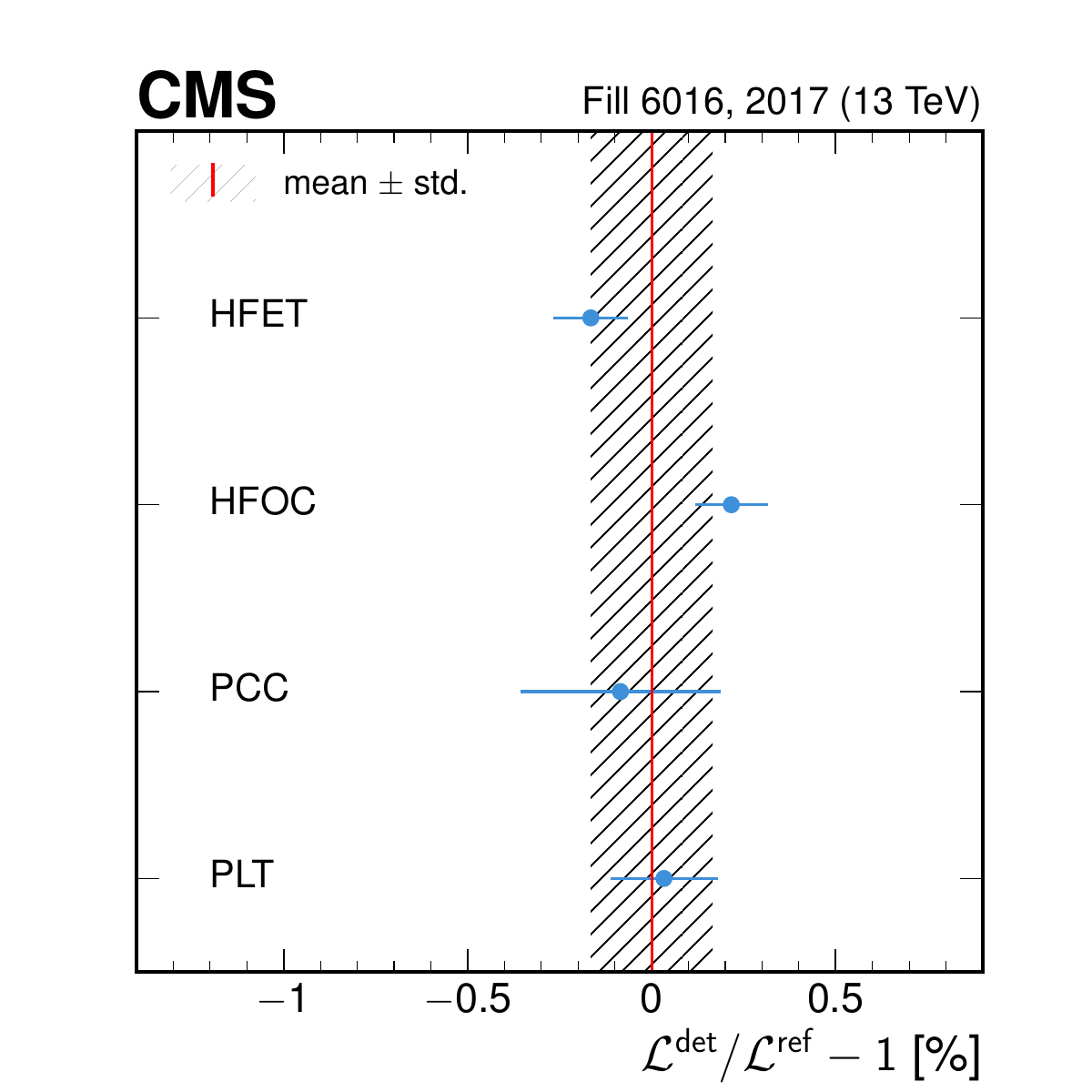}
\includegraphics[width=0.475\textwidth]{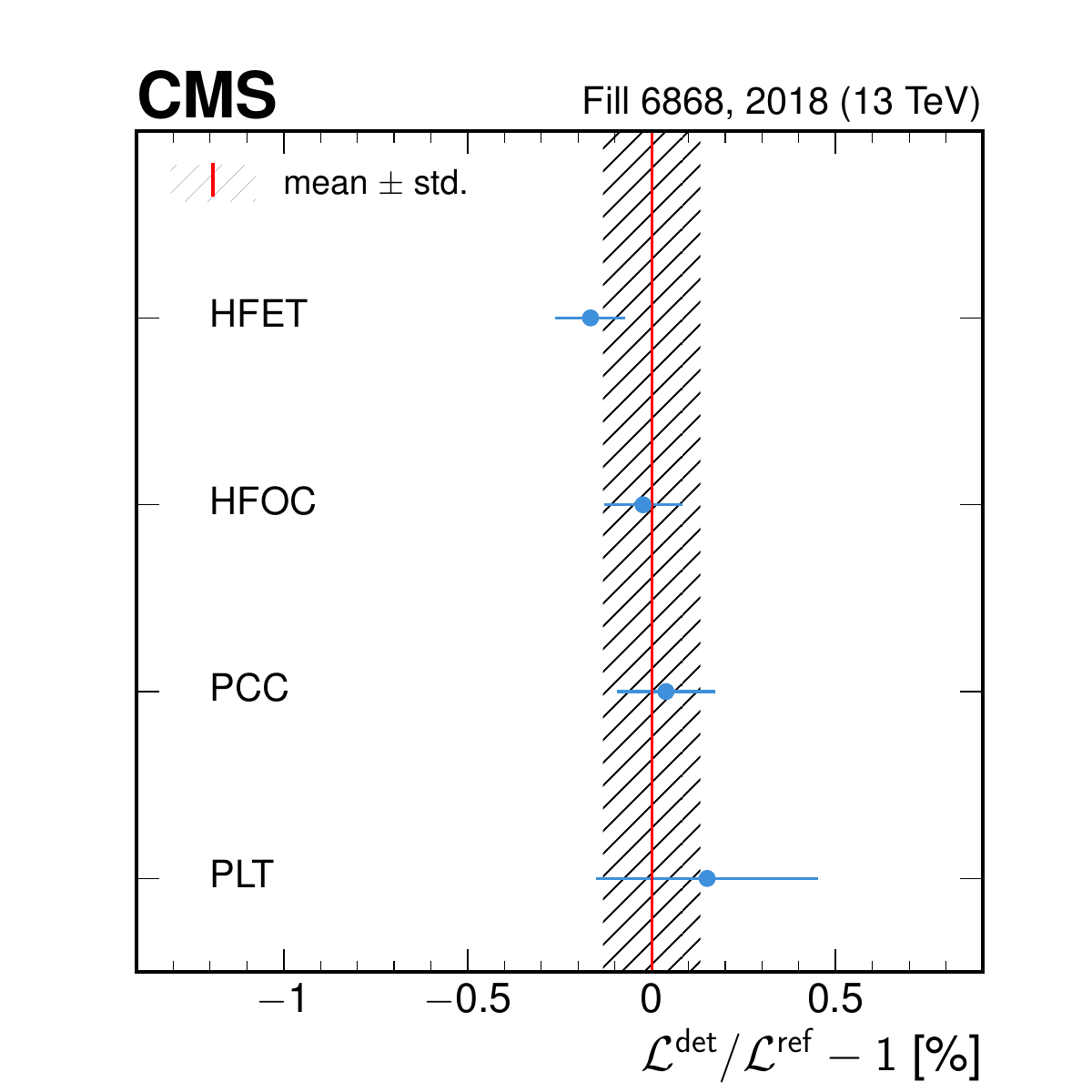}
\caption{
    Relative differences of the luminosity measured by the HFET, HFOC, PLT, and PCC detectors to the detector-averaged luminosity in the \vdM fill for the head-on periods for 2017 (left) and 2018 (right).
    The markers represent the average of ratios computed in approximately three-minute intervals, while the error bars indicate the standard deviation of the ratio distribution.
    The shaded area corresponds to the standard deviation of the individual averages.
}
\label{fig:vdM_xdet2}
\end{figure*}

Figure~\ref{fig:vdM_xdet2} illustrates the calibration consistency in the \vdM fills by showing, for each detector, the average ratio of its measured luminosity to the detector-averaged luminosity, computed in approximately three-minute intervals during periods with head-on collisions.
The shaded area represents the standard deviation of the individual ratio averages
and is taken as the uncertainty associated with the
\vdM calibration consistency. This amounts to 0.17 and 0.14\% in 2017 and 2018, respectively.

\section{Rate corrections for the high-luminosity regime}
\label{sec:detector_corrections}

The absolute calibration of each luminometer is determined using special scans that involve a small number of distinctly spaced proton bunches, characterized by high emittance, low bunch intensity, and low \instlumi.
Conversely, during regular operations, luminosity rates are recorded with nonisolated proton bunches packed into trains, under conditions of high bunch intensity and instantaneous luminosity, aimed at generating large data sets for physics measurements.
To ensure the consistent application of the absolute luminosity calibration, the measured rates must be corrected for various effects that become significant
under these data-taking conditions.
Such effects encompass out-of-time contributions,  nonlinear response, and detector efficiency changes.

Out-of-time (OOT) refers to rate contributions that are not the direct result of the collisions in the given 25\unit{ns} bunch crossing time window.
These contributions arise either from signal spillover from the preceding colliding bunch pair (Type 1) or from afterglow due to activation of detector material,  characterized by an exponentially decaying profile (Type 2).
Furthermore, pedestal levels due to instrumental noise or other sources also need to be accounted for.
The luminometer response can deviate from a linear relationship with \instlumi due to factors such as combinatorial or thresholding effects.
The efficiency of the detectors may change with operating conditions, such as high-voltage settings, and degrade over time due to radiation damage.

These effects vary in magnitude across the different luminometers.
In the following, the rate corrections specific to each detector are described prior to the application of the absolute calibration derived from \vdM scans.
The following conventions apply: a nonlinearity value of 1\%/(Hz/$\mu$b) indicates 
that the measured luminosity is overestimated by 1\% for each unit of single-bunch instantaneous luminosity (SBIL), when expressed in Hz/$\mu$b. 
Similarly, an efficiency value of 1.01 indicates that, at a given SBIL, the measured luminosity is overestimated by 1\%.

Efficiency and linearity are intrinsic detector properties, and the corresponding corrections are derived independently for each detector using short vdM-like ``emittance scans'' performed during regular data-taking conditions in each fill.
These scans follow a similar procedure to regular vdM scans but typically involve fewer points (7 in 2017, or 9 in 2018) and shorter durations (10 seconds per step).
The absolute calibration precision from these scans is limited by several factors, such as long-range electromagnetic interactions altering the bunch separation during scans, and a lack of a handle on the factorization effects. 
Therefore, $\sigmaVis^{\text{emit}}$ values from emittance scans are most comparable when taken in similar beam conditions and are treated as relative rather than absolute measurements, normalized to the value obtained in a regular fill near the vdM calibration program.

During 2017--2018, two scans per fill were typically performed, one at the start and another shortly before the end of the fill, with peak SBIL values that typically differ by a factor of two or more at the start.
Comparing $\sigmaVis^{\text{emit}}$ values as a function of SBIL allows for the determination of the extent of nonlinearity.
Studying the $\sigmaVis^{\text{emit}}$ values obtained over time, 
the changes in the detection efficiency can be observed as a function of integrated luminosity.

\subsection{Pixel luminosity telescope}
\label{sec:plt}

PLT rate corrections address nonlinearity and efficiency stability, as its signal is captured within the 25\unit{ns} of a single bunch crossing without being affected by OOT effects.
Detailed insights into the PLT luminosity measurement can be found in Ref.~\cite{CMSBRIL:2022qin}.
The results presented in this paper build on the telescope selection criteria and on the stability and linearity corrections derived solely from emittance scan data, as described in Ref.~\cite{CMSBRIL:2022qin}.
The 2017 luminosity exhibits residual effects, as the internal linearity and stability corrections derived from emittance scans rely on a limited number of data points. The efficiency is corrected using the emittance scan analysis method~\cite{CMSBRIL:2022qin}, which provides a time-dependent correction ranging from $-0.27$ to $+1.21\%$. 
The nonlinearity correction, derived relative to HFET as a function of SBIL, varies between 0.05 and 1.15\%/(Hz/$\mu$b). 
After applying these corrections, the luminosity measurements from PLT and HF remain uncorrelated.

\subsection{Hadron forward calorimeter}
\label{sec:hf}

Both HFET and HFOC suffer from OOT effects.
The signal spillover and the afterglow are quantified using data from fills with well-separated, solitary bunches.
Corrections, using the derived single bunch response function, are applied to the raw rates.
This methodology, inherited from prior publications and elaborated on in Ref.~\cite{CMS:2021xjt}, typically shows the largest impact on the bunch crossing immediately following a colliding bunch.
The integrated correction over the full orbit from spillover and material activation is typically 12\% for HFET and 16.5\% for HFOC, depending on the filling scheme.
The amount of spillover has a complex dependence on SBIL and introduces a small nonlinearity of $0.2\%/(\text{Hz}/\mu$b) for both counting methods,
even after applying OOT corrections that are assumed to be linear in SBIL.
For HFOC, which employs a threshold when counting the calorimeter towers with energy deposition, an additional ``bunch train effect''~\cite{CMS:2021xjt} arises when
the OOT contribution and the signal from genuine collisions are individually below the threshold, but together exceed it.
A correction is derived by studying the ratio of the first-in-train HFOC bunch measurement over the average of the next four bunches divided by the same ratio for HFET.
The resulting double ratio is fitted as a function of SBIL with a second-order polynomial, yielding an additional nonlinearity correction of 0.2\%/(Hz/$\mu$b).

\begin{figure}[!htpb]
\centering
\ifthenelse{\boolean{cms@external}}{
\includegraphics[width=0.45\textwidth]{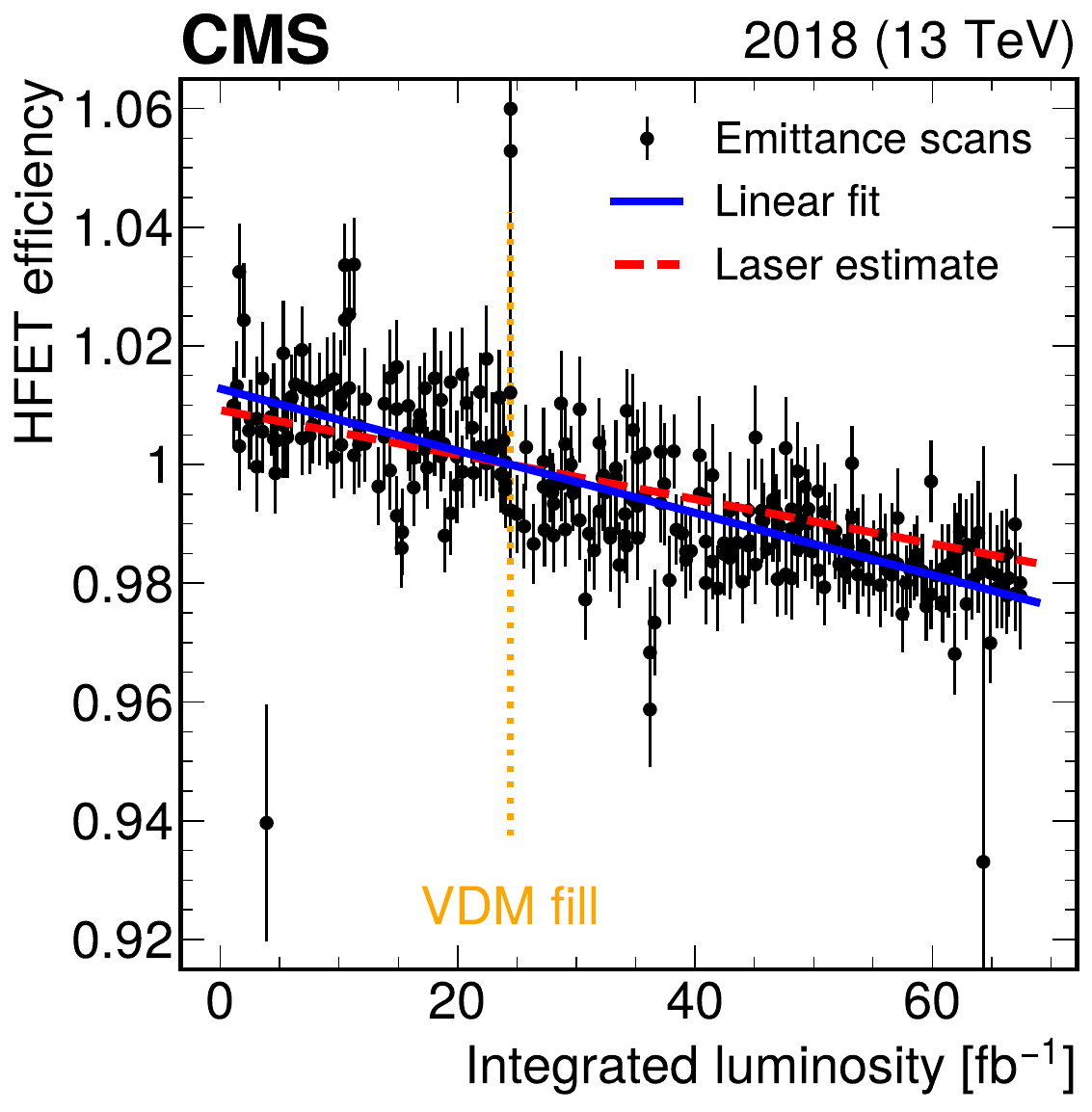}
}{
\includegraphics[width=0.65\textwidth]{Figure_012.pdf}
}

\caption{%
    The decrease in efficiency as a consequence of HFET aging in 2018 as measured using emittance scans.
    The black points represent the efficiency of the HFET counting method relative to the efficiency at the time of the vdM scan (indicated by the vertical orange line), as a function of the total integrated luminosity in 2018. Significant outliers were examined for potential fit quality problems, however, none were identified. 
    The uncertainties shown are the standard deviations computed over the BCIDs.
    The blue line is a linear fit to the data.
    The expected aging from laser calibration monitoring of the fibers and photomultipliers is indicated by a red line and shows a good agreement with the emittance scan data.
}
\label{fig:hfaging}
\end{figure}

The aging of photomultipliers and optical fibers in the HF calorimeters leads to a gradual reduction in efficiency for both HFOC and HFET.
To ensure stable luminosity measurements over time, the emittance scan analysis method is employed.
The resulting correction depends on the integrated luminosity delivered to CMS, amounting to 
0.05\%\fbinv for HFET, and 0.03\%\fbinv for HFOC.
Figure~\ref{fig:hfaging} shows the measured relative efficiency of HFET in 2018 data.
As a validation, an independent estimate of the efficiency is obtained using the HF laser system signal. The two methods show good agreement.

\subsection{Pixel cluster counting}
\label{sec:pcc}

A stringent selection is applied to the pixel modules that enter the luminosity measurement.
The criteria involve the noise in the vdM calibration fill and the response stability and collinearity during the full year of data taking evaluated relative to the average response of the selected modules. The module selection is performed in several iterations, using progressively tighter selections and the increasingly stable average response of the modules selected in the previous iteration as the reference. In all cases, the innermost barrel layer is excluded due to the high occupancy in this region. 
In 2017 PCC data, the vdM rates for pixel modules in the barrel  are inconsistent with the vdM scan profiles of all other detectors. However, the forward pixel module rates show a consistent scan shape with all other luminometers. As a result, only the forward pixel modules are used for vdM calibration and luminosity measurements in the 2017 data.

\begin{figure}[!htpb]
\centering
\ifthenelse{\boolean{cms@external}}{
\includegraphics[width=0.45\textwidth]{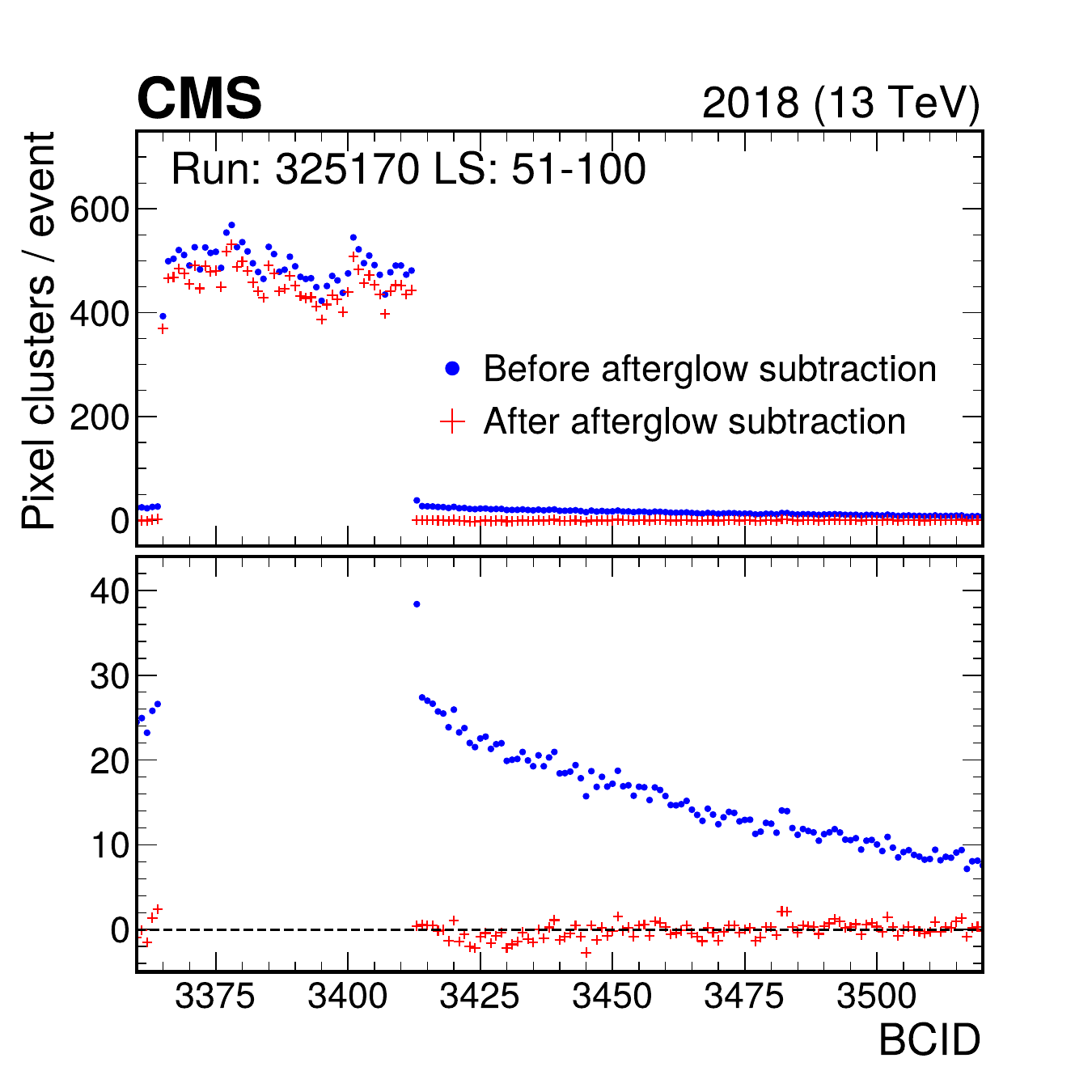}
}{
\includegraphics[width=0.65\textwidth]{Figure_013.pdf}
}
\caption{
    The number of clusters in the pixel tracker per event is shown as a function of the BCID for the last bunch train of the orbit. Blue points represent the values before correction, while red crosses indicate the results after applying out-of-time corrections.
    The upper panel displays the full count range, encompassing both colliding and empty BCIDs, while the lower panel focuses on empty bunch crossings with a different scale for improved visibility.
    In the lower panel, the red crosses, representing the residual rate, are close to zero, demonstrating the excellent performance of the correction.
}
\label{fig:pcc_afterglow}
\end{figure}

Pixel cluster counting is also subject to OOT effects.
The corresponding contributions are determined using random trigger data averaged over periods of about 20 minutes.
The rates in each time block are fit as a function of BCIDs using a model that includes both Type\,1 and Type\,2 OOT components for each colliding bunch as described in Ref.~\cite{CMS:2021xjt}.
The exponentially decaying afterglow component uses two parameters, one for normalization relative to the rate from the colliding bunch and one for the decay constant.
These parameters are predetermined in representative fills for the year.
The Type\,1 component is time dependent and is left floating in the fit.
The total OOT background, integrated over the full orbit, varies between 5 and 8\% for typical LHC fills in physics conditions.
Figure~\ref{fig:pcc_afterglow} shows per-bunch pixel cluster counts per event from a 2018 fill before and after the application of the OOT corrections in the vicinity of the last bunch train. 
The increased contribution from spillover is clearly visible for the first bunch after the train on top of the exponential activation tail.

An analysis of the emittance scan data is not possible due to the low recording rate of the zero bias data set in physics conditions.
Although the linearity of the PCC method has been studied through simulations of the detector~\cite{CMS:2021xjt}, there is no correction for the expected sub-percent nonlinearity.

The PCC suffers from dynamic inefficiency due to the limited bandwidth of the readout system and the finite capacity of the front-end buffers.
The loss of efficiency depends on the orbit-integrated \instlumi and has been quantified by analyzing missing hits along reconstructed tracks using the method described in Ref.~\cite{TRK-11-001} in 2017 and 2018 separately.
Considering the module selection, the impact on luminosity is determined to be 0.030\%/(Hz/nb) and 0.018\%/(Hz/nb) in 2017 and 2018, respectively, and is applied as a correction, changing the total integrated luminosity by 0.31\% in 2017 and 0.22\% in 2018 as measured by PCC.

\section{Luminosity integration}
\label{sec:luminosity_integration_and_uncertainty}

The luminosity integration entails applying, for a full year of physics operations, the absolute vdM calibration to the luminometer rates corrected as described in Section~\ref{sec:detector_corrections}.
The CMS experiment relies on multiple independently calibrated systems (HFET, HFOC, PLT, and PCC in 2017--2018), and the final luminosity is derived from averaging these measurements (``mean luminosity'').
Differences in the measured luminosity among the luminometers arise from statistical variations, the limited precision of the absolute calibrations and of the applied rate corrections, as well as uncorrected instrumental effects.

To estimate the systematic uncertainty due to residual nonlinearity, we adopt the methodology of previous publications~\cite{CMS:2021xjt,CMS-PAS-LUM-17-004, CMS-PAS-LUM-18-002}.
In each fill, the gradually decreasing luminosity provides an opportunity to examine the relative nonlinearity of the mean luminosity with respect to a reference detector.
A linear fit to the ratio of the instantaneous luminosities, as measured by the mean luminosity method and the reference luminometer, as a function of the reference luminometer SBIL is performed.
The reference luminometer is either DT or RAMSES because of their robust linear response~\cite{CMS:2021xjt}.
Only fills with an SBIL range spanning at least 2\unit{Hz/\(\mu\)b} and a duration of at  least 30 minutes are considered. The luminosity information is averaged over approximately 6-minute blocks for the ratio calculation.

\begin{figure}[!htpb]
\centering
\includegraphics[width=0.45\textwidth]{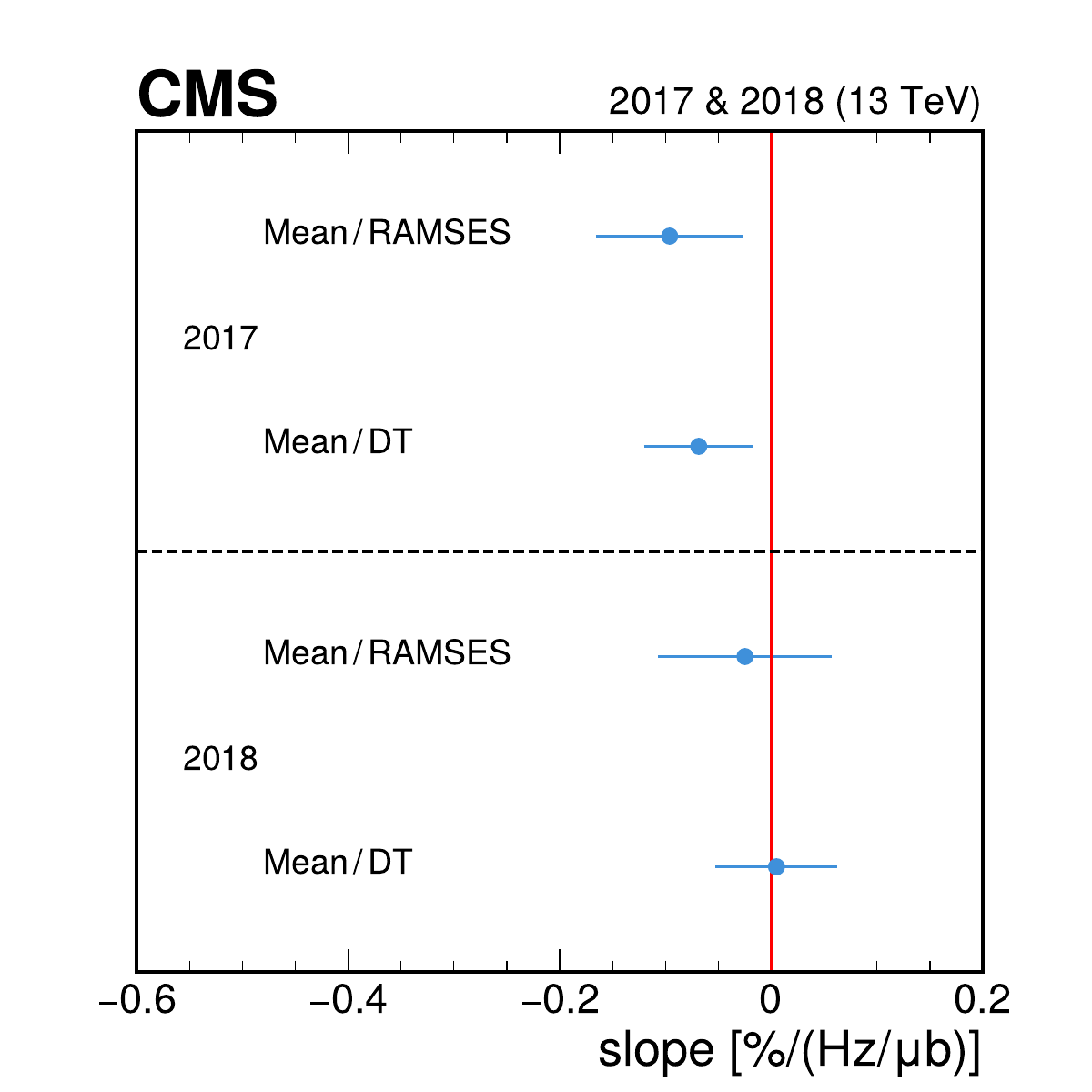}
\caption{
    The luminosity-weighted average of the measured residual nonlinearity between the mean luminosity and DT or RAMSES for 2017 and 2018.
    The error bars signify the weighted standard deviation over the individual fills.
}
\label{fig:stabilitylinearity:linearity3}
\end{figure}

Figure~\ref{fig:stabilitylinearity:linearity3} shows the luminosity-weighted mean and standard deviation of the slopes obtained by comparing the mean luminosity with either DT or RAMSES.
As the basis for the uncertainty calculation, the largest among the two mean values and the two standard deviations is used, evaluated separately for each year.
Accordingly, for 2017 the mean value, and for 2018 the standard deviation of the slopes with respect to RAMSES are used to estimate the uncertainty, yielding 0.095\%/(Hz/\(\mu\)b), and 0.082\%/(Hz/\(\mu\)b) respectively.
The final linearity uncertainty of the integrated luminosity is obtained by multiplying these slope values by the luminosity-weighted mean SBIL of 5.34\unit{Hz/\(\mu\)b} (5.12\unit{Hz/\(\mu\)b}), resulting in a linearity uncertainty of 0.51\% (0.42\%) for 2017 (2018).

\begin{figure*}[!htpb]
\centering
\includegraphics[width=0.45\textwidth]{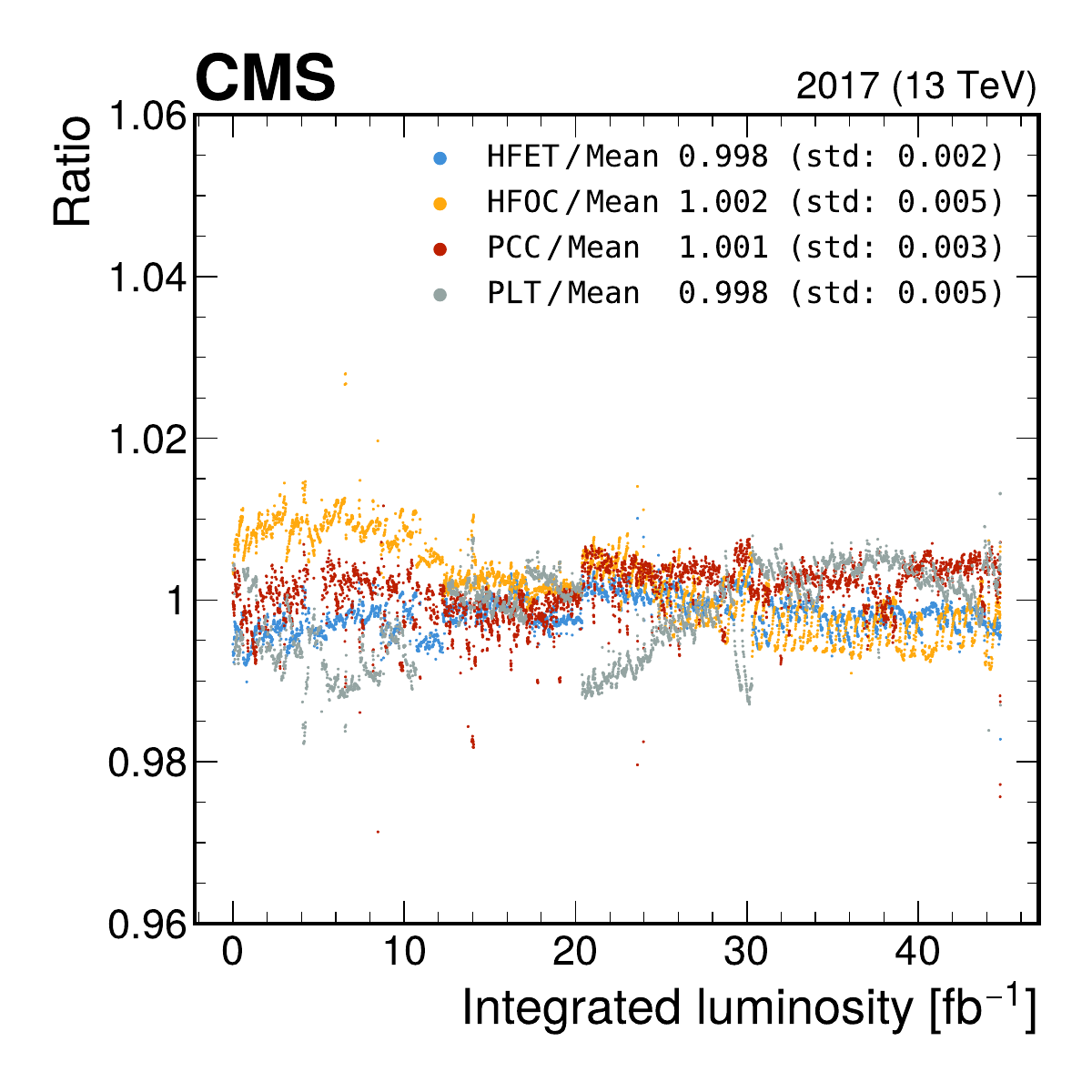}
\includegraphics[width=0.45\textwidth]{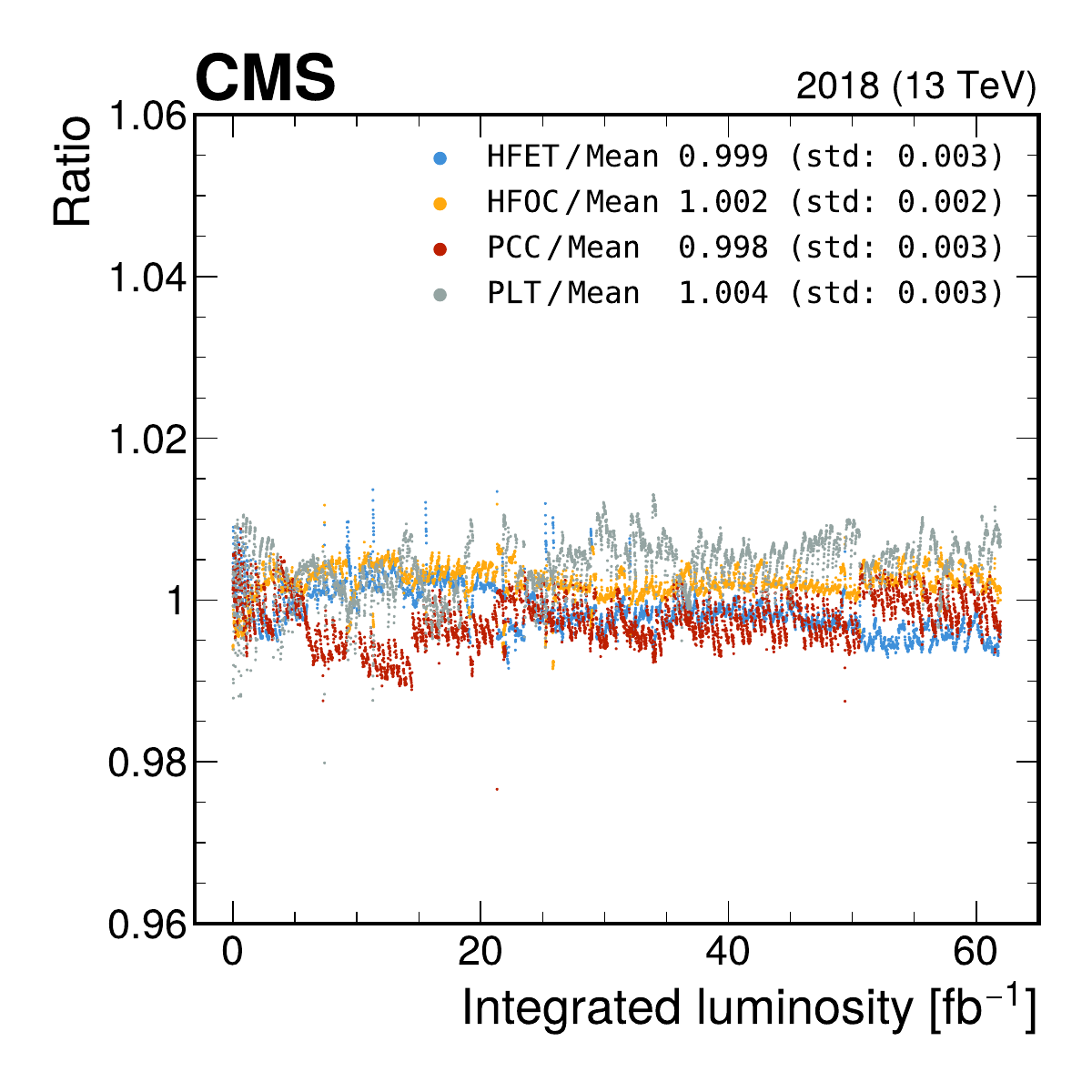} \\
\includegraphics[width=0.45\textwidth]{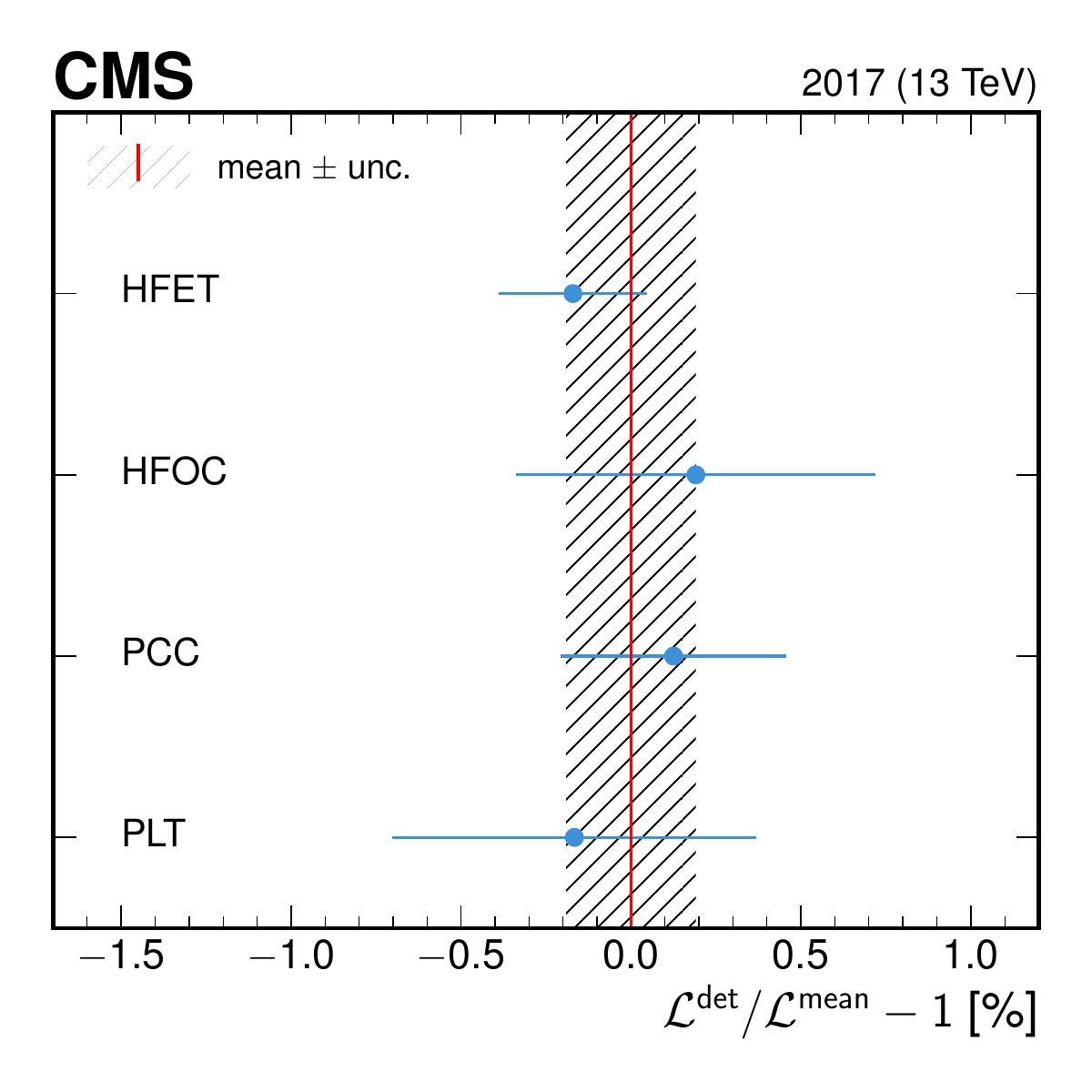}
\includegraphics[width=0.45\textwidth]{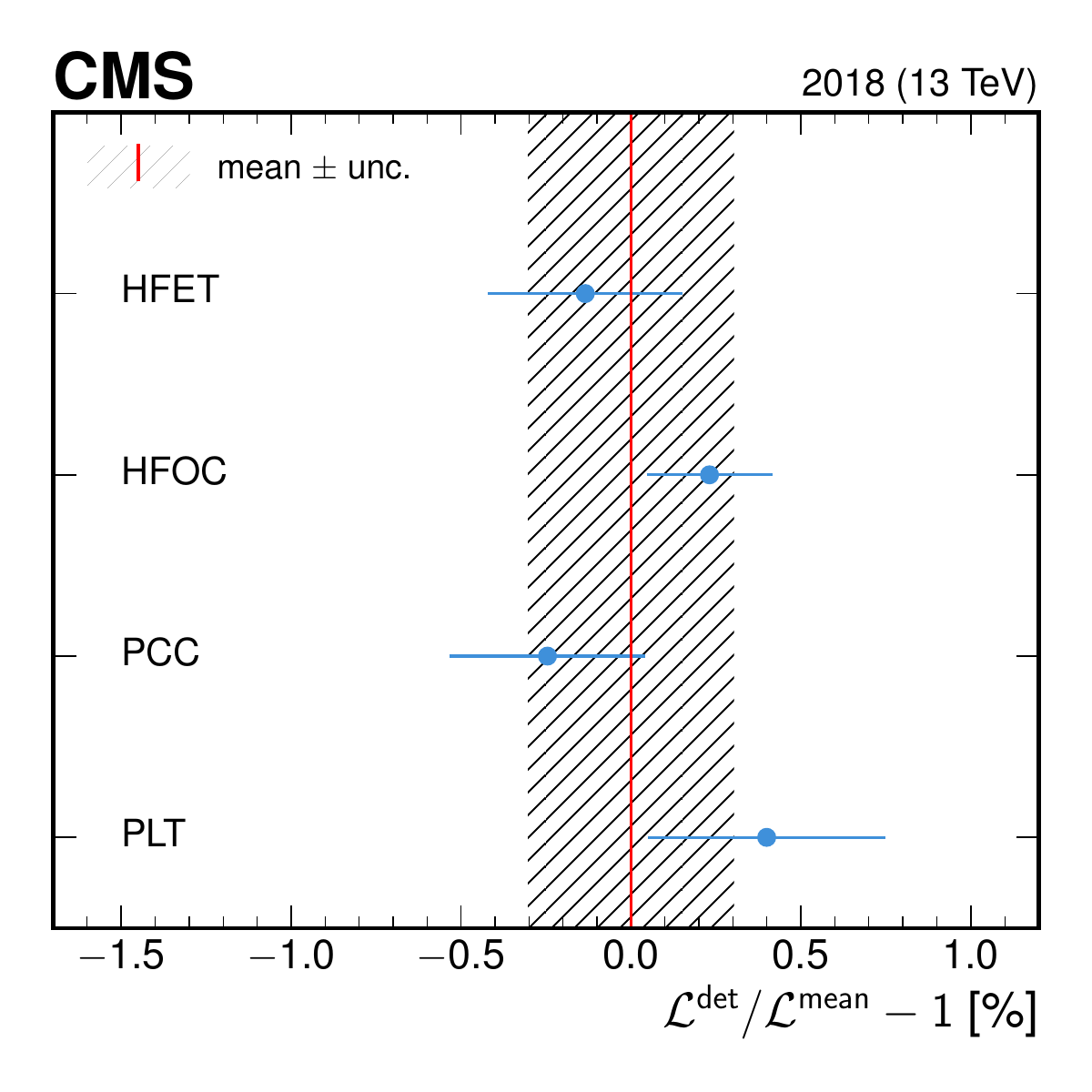}
\caption{
    The ratio of the luminosity measured by HFET, HFOC, PLT, and PCC to their mean per 20-minute blocks for 2017 (left) and 2018 (right).
    The upper row shows the ratio as a function of integrated luminosity, while the lower row displays the luminosity-weighted average of the ratios.
    The error bars signify the weighted standard deviation over the 20-minute units.
    The striped area represents the unweighted standard deviation of the four average ratio values.
}
\label{fig:stabilitylinearity:stability}
\end{figure*}

A systematic uncertainty that reflects the consistency of the individual measurements with the mean luminosity is assessed.  Figure~\ref{fig:stabilitylinearity:stability} shows the ratio of the individual luminosity values to the mean luminosity.
The plots in the upper row present these ratios as functions of the integrated luminosity, while the lower row compares their luminosity-weighted averages.
We take the standard deviation of the four average ratio values to be the final consistency uncertainty, amounting to 0.19 and 0.30\% for 2017 and 2018, respectively.

A similar evaluation is performed for the low-pileup (low-PU) data set, which represents a distinct running regime compared to the standard high-pileup (high-PU) data taking considered up to this point.
These data were recorded in a handful of fills near the end of the 2017 data taking, with an average SBIL of only 0.41\unit{Hz/\(\mu\)b}. This corresponds to a pileup of approximately 3.
In this regime, PCC is not available for luminosity measurement due to low cluster counts.
The sample standard deviation of the remaining three luminosity sources is 0.67\%, which is similar to the spread exhibited by these detectors in the regular high-pileup conditions by the end of the year.

The linearity uncertainty is estimated using the relative nonlinearity derived in high-PU conditions for the mean of HFOC, HFET, and PLT with respect to DT and RAMSES.
The value used (0.113\%/(Hz/\(\mu\)b)) is the average relative slope between the three-detector mean and RAMSES.
Multiplied by the mean SBIL in the low-PU era results in an uncertainty of 0.04\%.

\subsection{Consistency with the \texorpdfstring{\PZ}{Z} boson rate measurements}
\label{sec:zcounting}

Simple luminometer-specific signals such as (coincidence) counts and deposited energy sums are not the only possible proxies for luminosity.
The rate of any particle physics process with sufficient statistical power, a clean signature, and a practical way to track the reconstruction efficiency under the constantly changing operating conditions can be set to this purpose.

The production rate of \PZ bosons that subsequently decay into a pair of muons was originally proposed 
as a handle on luminosity~\cite{Salfeld-Nebgen:2018bdp} and was first demonstrated at CMS in Ref.~\cite{CMS:2024zco}. In this paper, this method is extended to 
2016--2018 to validate the stability of the luminosity measurement across these years.

The luminosity for a data set with $N_{\text{obs}}$ selected events and $N_{\text{bkg}}$ expected background events is given by:
\begin{linenomath}
\begin{equation}
    \intlumi = \frac{N_{\text{obs}}-N_{\text{bkg}}}{\sigZ\,\AZmm\,\effZmm} = \frac{\NZ}{\sigZfid\,\effZmm},
\end{equation}
\end{linenomath}
where \sigZ is the known \pp production cross section of \PZ bosons decaying into a pair of muons, \AZmm is the detector acceptance defined at the generator level, and \effZmm is the time-dependent experimental selection efficiency.
The product $\sigZfid=\sigZ\,\AZmm$ is constant for a given center-of-mass energy but is not known to a sufficient accuracy due to the considerable theoretical uncertainties primarily related to the parton distribution functions.
Nonetheless, the $\PZ\to\MM$ rate can be treated as a relative measurement.

The \PZ boson rate is measured in intervals of 20\pbinv, each spanning about 20 minutes, using an in-situ determination of the most relevant effects.
The background contribution is subtracted by fitting signal and background templates to the invariant mass of the dimuon system.
The full muon efficiency is tracked via tag-and-probe measurements in each interval.
The independently reconstructed muon tracks in the silicon tracker and muon chamber are used to measure the full track efficiency.
Small residual corrections at the per-mille level are taken into account based on simulation. These  arise from correlations of the single muon trigger selection between the two muons, from correlations between the silicon tracker and muon chamber tracks of each muon, and the effects of detector resolution on the measured muon momentum.

\begin{figure}[!htb]
\centering
\ifthenelse{\boolean{cms@external}}{
\includegraphics[width=0.45\textwidth]{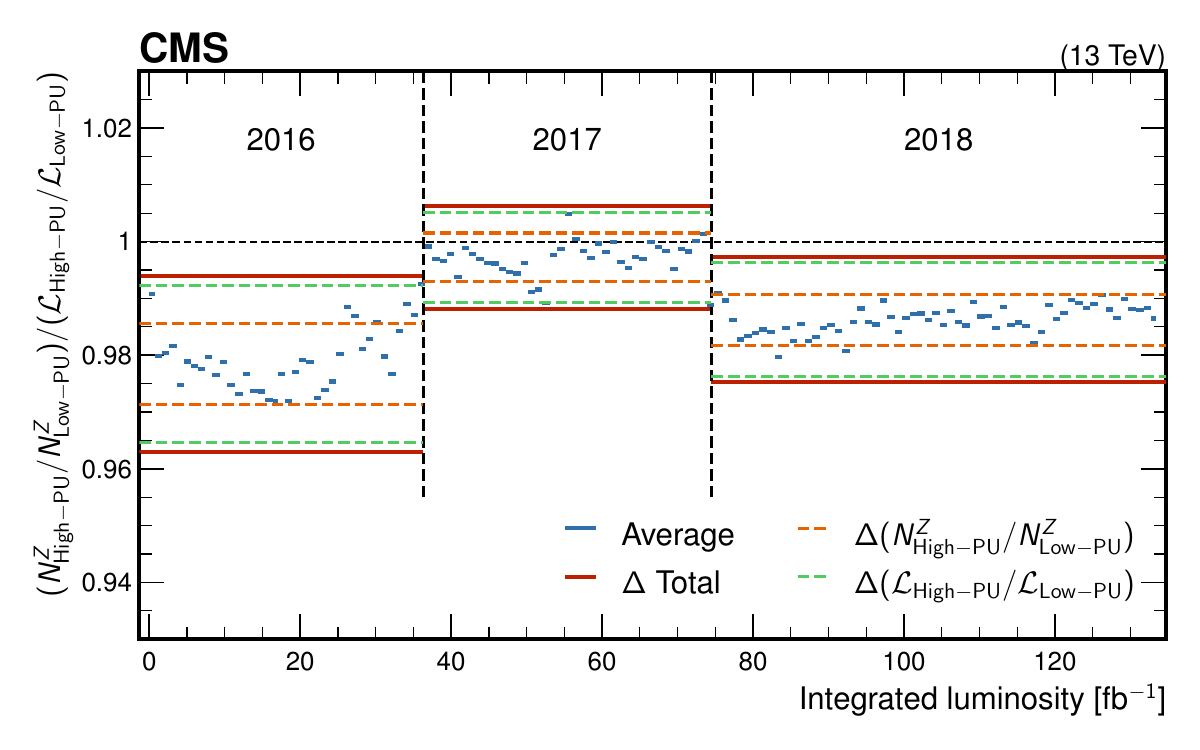}
}{
\includegraphics[width=0.65\textwidth]{Figure_016.pdf}
}

\caption{
    The luminosity as determined by \PZ boson production rates divided by the reference luminosity as a function of the reference integrated luminosity for the 2016--2018 high-PU data set.
    Each blue line section corresponds to the luminosity values measured in about 1\fbinv of data, used for the measurement of $\NZ$.
    The green and orange dashed lines indicate the uncertainties from the \NZ ratio and reference luminosity ratios, respectively, for each year of data taking.
    Accordingly, the red lines indicate the total uncertainty in the double ratio.
    The vertical dashed lines separate the 2016, 2017, and 2018 data.
}
\label{fig:zcounting_stability_scatter}
\end{figure}

A qualitative assessment of the luminosity in the high-PU regions is performed by
anchoring \sigZfid in the low-PU era of 2017:
\begin{equation}
    \lumiZ = \frac{\NZ_\highPU}{\sigZfid} = \frac{\intlumi_\lowPU}{\NZ_\lowPU}  \NZ_\highPU.
\label{eq:NZ_lumitransfer}
\end{equation}
The uncertainty in the ratio of \PZ bosons $\NZ_\highPU/\NZ_\lowPU$ was estimated to be 0.43\% in 2017~\cite{CMS:2024zco}.
From a similar study for ratios of the 2016 and 2018 high-PU to 2017 low-PU \PZ bosons counts, the uncertainties are 0.72 and 0.45\%, respectively.
As seen in Fig.~\ref{fig:zcounting_stability_scatter}, a good stability is observed for each year independently between the \PZ boson luminosity and the luminometer-based measurement reported above (and in Ref.~\cite{CMS:2021xjt} for 2016) based on the CMS bunch-by-bunch capable luminometers (called reference luminosity hereafter), with steps in the ratio of about 1.5\% between 2016 and 2017 and about 1\% between 2017 and 2018.
These values can be attributed to the different vdM calibrations used to normalize the reference luminosity in 2016 and 2018, but also in part to effects in the \PZ boson rate measurements, which differ with the year of data taking.
In the middle of 2017, a step of 0.5--1\% is observed, identified with a similar shift in the PLT measurement.

To assess the compatibility of the measurements based on the luminometers and on the \PZ boson production rate, a likelihood combination is performed taking into account all uncertainties and their correlations.
The likelihood ($L$) includes $\chi^2$ terms for the number of predicted ($\NZ_{i, \text{pred}}$) and extracted ($\NZ_{i}$) \PZ bosons for each data set, with $\sigma_i$ as the statistical uncertainty of $\NZ_{i}$, and a set of Gaussian-constrained nuisance parameters ($\theta_j$) to incorporate systematic uncertainties:
\begin{equation}
    -2\ln(L) = \sum_i^\text{data set} \frac{1}{\sigma_i^2} 
    \Big(N_{i, \text{pred}}^\PZ(\vec{\theta}) - N_{i}^\PZ\Big)^2 + \sum_j^\text{syst} \theta_j^2.
\end{equation}
The number of predicted \PZ bosons is given by
\begin{equation}
    \NZ_{i,\mathrm{pred}} = \intlumi_i\big(\vec{\theta}\big) \effZmm_i\big(\vec{\theta}\big) \sigZfid,
\end{equation}
with the luminosity ($\intlumi$) and full \PZ boson reconstruction and identification efficiency (\effZmm) that depend on a set of nuisance parameters.
The fiducial cross section (\sigZfid) is treated as an unconstrained parameter of interest.

\begin{figure}[!htpb]
\centering
\ifthenelse{\boolean{cms@external}}{
\includegraphics[width=0.45\textwidth]{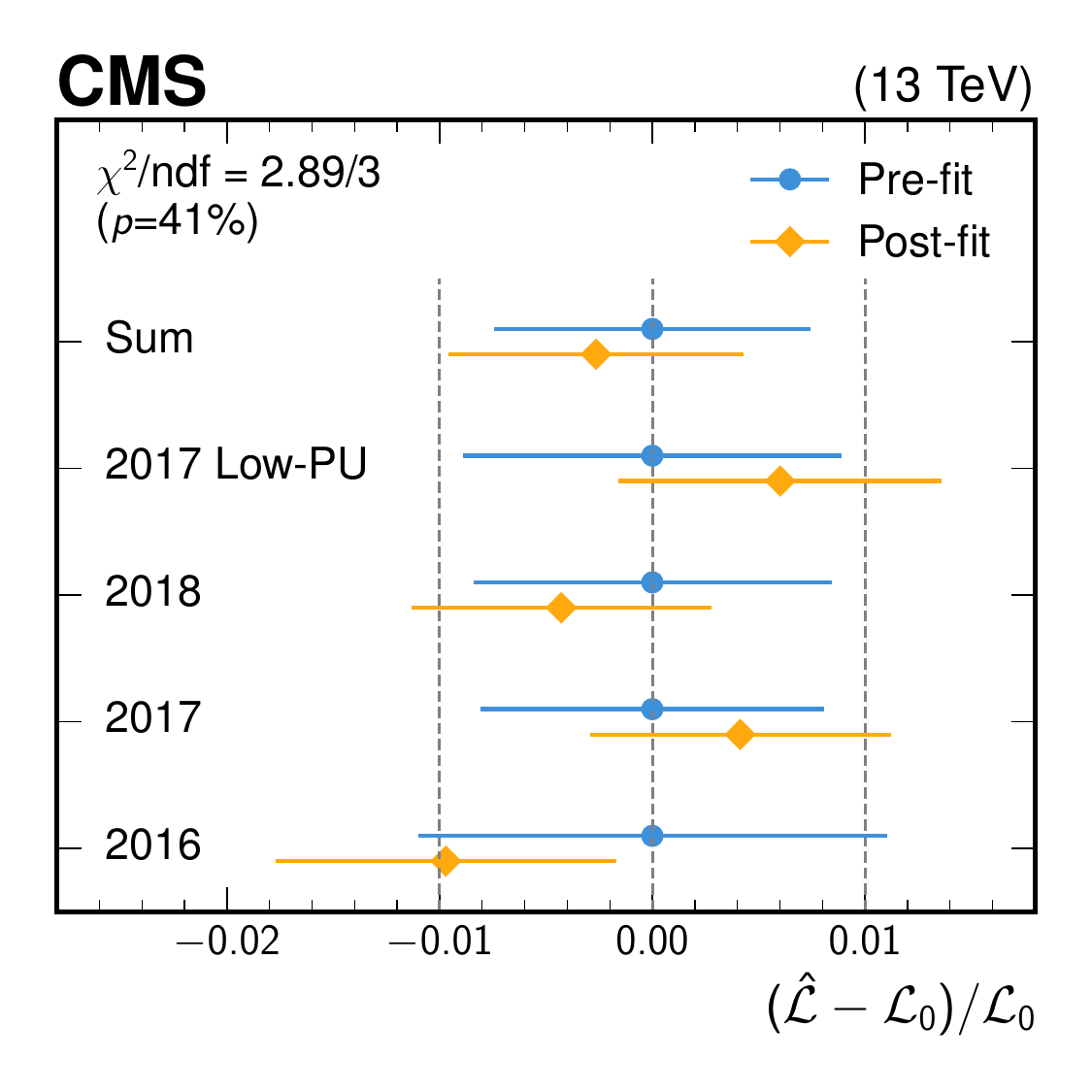}
}{
\includegraphics[width=0.65\textwidth]{Figure_017.pdf}
}

\caption{
    The relative uncertainty and the shifts of the luminosity values before and after the likelihood fit.
    The pre-fit (blue) results originate from the traditional luminosity estimation using the luminometers, while the post-fit (orange) values also include the \PZ boson rate data.
    The figure also shows the reduced $\chi^2$ and the $p$-value of the fit.
}
\label{fig:zcounting_combination_lumi}
\end{figure}

\begin{figure}[!htpb]
\centering
\ifthenelse{\boolean{cms@external}}{
\includegraphics[width=0.45\textwidth]{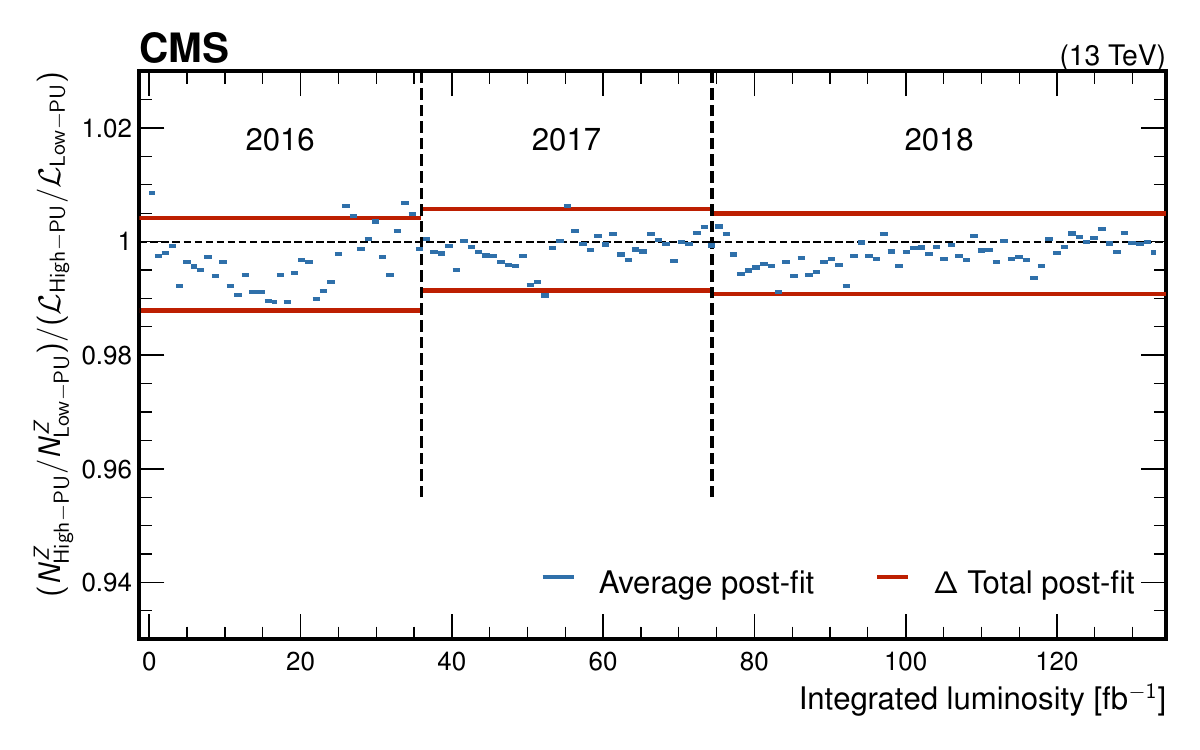}
}{
\includegraphics[width=0.9\textwidth]{Figure_018.pdf}
}\caption{
    The post-fit ratio of the luminosity based on measured \PZ boson production rates and the reference luminosity as a function of the reference integrated luminosity for the 2016--2018 high-PU data set.
    Each blue line section corresponds to the luminosity values measured in about 1\fbinv of data.      The red lines indicate the total uncertainty in the double ratio.
    The vertical dashed lines separate the 2016, 2017, and 2018 data.
}
\label{fig:zcounting_stability_scatter2}
\end{figure}

The result of minimizing the negative log-likelihood function including all data sets is shown in Figs.~\ref{fig:zcounting_combination_lumi}--\ref{fig:zcounting_stability_scatter2}.
The latter figure shows how the ratio in all three years is closer to unity than in  Fig.~\ref{fig:zcounting_stability_scatter}.
Examining the pulls on the uncertainty sources in the combination, the 2017 vdM scan-to-scan nuisance parameter is consistent with the mean within 0.6 standard deviations, while the 2018 integrated cross-detector stability is pulled down by 0.6 standard deviations.
All other nuisance parameters remain within 0.5 standard deviations.
In summary, all measurements are found to be consistent within one standard deviation.
The measured cross section $\sigZfid=708\pm 12\unit{pb}$, is in agreement with the theoretical prediction of $734\pm36\unit{pb}$~\cite{cuteMCFM}.
To assess the goodness-of-fit, a saturated likelihood test is performed, leading to a $p$-value of 41\%, therefore, no tension between the two measurements has been found.
The final uncertainties and luminosity values reported in the summary do not incorporate the result of this fit.

\section{Results}
\label{sec:results}

The methods used to determine the integrated luminosity for the 13\TeV proton-proton data collected at CMS in 2017--2018, as well as the sources of systematic uncertainties limiting its precision, have been thoroughly discussed above.
In Table~\ref{TAB:SystematicError}, these sources are divided into two categories: ``normalization'' uncertainties, which impact the absolute luminosity scale determined using the vdM scan method, and ``integration'' uncertainties, which include 
linearity of the detector response, as well as the final consistency of the results among the independently calibrated systems.

\begin{table*}[!ht]
\centering
\topcaption{
    Summary of the luminosity uncertainties in 2017 and 2018, divided into two groups affecting the \vdM calibration at low luminosity (normalization), and the measurement of
    the luminosity in physics conditions (integration).
    The ``Corr.'' column indicates whether the uncertainties are fully positively correlated between the two years or independent, while the last column lists the presumed correlation with the 2015--2016 data set.
}
\label{TAB:SystematicError}
\renewcommand{\arraystretch}{1.1}
\cmsTable{
\begin{scotch}{llccccc}
    
    \multicolumn{2}{l}{Source} & 2017 [\%] & 2018 [\%] & Corr. & Corr. w. 15--16  \\
    \hline
    \multicolumn{2}{l}{{Normalization}} & & & & & \\
    & \hspace{+2mm} Luminometer backgrounds & 0.07 & 0.07 & No & \NA\\
    & \hspace{+2mm} Beam current normalization & 0.20 & 0.20 & Yes & Yes\\
    & \hspace{+2mm} Ghost and satellite charges & 0.06 & 0.07 & Yes & Yes\\
    & \hspace{+2mm} Beam-beam effects & 0.29 & 0.30 & Yes &  Yes\\
    & \hspace{+2mm} Orbit drift & 0.09 & 0.19 & Yes & Yes\\
    & \hspace{+2mm} Length scale calibration & 0.15 & 0.18 & Yes & No\\
    & \hspace{+2mm} Transverse factorizability & 0.29 & 0.33 & Yes & No\\
    & \hspace{+2mm} Scan-to-scan variation & 0.26 & 0.27 & No & No\\
    & \hspace{+2mm} Bunch-to-bunch variation & 0.10 & 0.10 & No & No\\
    & \hspace{+2mm} Cross-detector consistency at vdM & 0.17
    & 0.14 & No & No\\
    \multicolumn{2}{l}{{Integration}} & & & & \\
    & \hspace{+2mm} Cross-detector consistency per year & 0.19 
    & 0.30 & No & No\\
    & \hspace{+2mm} Linearity & 0.51 
    & 0.42 & Yes & No\\
    \multicolumn{2}{l}{Total normalization uncertainty} & 0.59 
    & 0.65 & \NA & \NA \\
    \multicolumn{2}{l}{Total integration uncertainty} & 0.54 
    & 0.52 & \NA & \NA \\
    \multicolumn{2}{l}{Total uncertainty} & 0.81 
    & 0.83 & \NA & \NA \\
    
\end{scotch}
}
\end{table*}

The primary sources of systematic uncertainties affecting the absolute calibration (Section~\ref{sec:VdMCalib}) stem from limitations in our understanding of beam-beam effects, the characterization of transverse factorizability of the bunch proton density in the transverse coordinates, and the variability in results across vdM scans performed under changing beam conditions during the calibration fill.
The dominant source of uncertainty in the measurement is residual nonlinearity (Section~\ref{sec:luminosity_integration_and_uncertainty}) for both years. 
This source may present challenges in the future as well, as it increases proportionally with the single bunch instantaneous luminosity for a fixed relative slope estimate.

Combining the normalization and integration uncertainties, which contribute similarly to the total for the high-PU data, the relative precision is 0.81\% in 2017 and 0.83\% in 2018.
The measured integrated luminosity with high-quality physics data 
amounts to $42.07\pm0.34\fbinv$ in 2017 and $59.56\pm 0.49\fbinv$ in 2018 in the high-PU periods.
In 2017, CMS also collected $0.2113 \pm 0.0019\fbinv$ of data at a pileup of approximately 3.

Normalization uncertainties are considered correlated across years when they arise from the same underlying physical processes or instrumental sources, or when the methodology dominates the uncertainty.
The only normalization uncertainties that are not correlated between the 2017 and 2018 vdM programs are due to the scan-to-scan and bunch-to-bunch variations of \sigmaVis, as well as the disagreement among the integrated luminosities measured in the vdM fill by the independently calibrated luminometers. These factors are partly statistical or random in nature, and their underlying source can vary from year to year. Additionally, the uncertainty from the background corrections for the individual luminometer data is treated independently as the two years employ different methodologies. 

Similarly, for the integration uncertainties, cross-detector luminosity comparisons are treated as uncorrelated,
whereas the uncertainty associated with linearity, which is assumed to be a more persistent quality of the detectors, is considered correlated.
The relative covariance matrix of the years is shown in Table~\ref{tab:rcorr}.
\begin{table}[ht!]
\centering
\topcaption{The relative covariance matrix of the luminosity measurements for high-pileup proton-proton collision data recorded by the CMS experiment during the Run 2 period, where each element is given as $\sigma_{i,j}/(\lumi_i \lumi_j)\times 10^4$. }\label{tab:rcorr}
\begin{scotch}{cc c c c }
{} & {2015} & {2016} & {2017} & {2018} \\
\hline
{2015} & 2.59 & 1.28 & 0.26 & 0.35 \\
{2016} & 1.28 & 1.52 & 0.24 & 0.29 \\
{2017} & 0.26 & 0.24 & 0.65 & 0.49 \\
{2018} & 0.35 & 0.29 & 0.49 & 0.69 \\
\end{scotch}
\end{table}

The combination of the total uncertainty values across different years reflects the improvement due to the uncorrelated nature of certain systematic effects and results in a relative precision of 0.73\% for both the full 2015--2018 and the restricted 2016--2018 high-pileup data sets.
The total integrated luminosity for high-quality high-pileup physics data collected during 2016--2018 (2015--2018) is $137.94\pm 1.00$ ($140.22\pm 1.02$)\fbinv.

\section{Summary}
\label{sec:summary}

In this work, we presented the legacy measurement of the integrated luminosity recorded by the Compact Muon Solenoid (CMS) detector in proton-proton collisions at a center-of-mass energy of
13\TeV during the Run~2 data-taking period (2015--2018) at the CERN Large Hadron Collider (LHC).
This measurement significantly improves upon previous CMS results, reducing the single-year uncertainties to the sub-percent level. This gain in precision was achieved through the introduction of new and refined methods to evaluate and correct several sources of bias.

These include effects related to micrometer-scale beam position monitoring, for which residual orbit drift corrections are introduced, as well as the calibration of the beam position length scale with respect to the CMS tracker coordinate system, for which a two-step approach is presented for the first time.
The impact of the factorization assumption for the bunch proton densities is addressed using a refined luminous region method which is also compared to an independent measurement based on two-dimensional luminometer rate fits. Improvements are introduced in the estimation of the uncertainties associated with beam-beam interactions.
In addition, meticulous corrections to the luminometer rates for out-of-time contributions, followed by the analysis of the emittance scan data to monitor detector performance, lead to a deeper understanding of the linearity and of the efficiency variations. 

The excellent agreement among the four independently calibrated and operated luminometers, using different detector technologies and counting methods,
provides a rigorous cross-check of the absolute measurement. By combining them for the first time, the overall uncertainty is further reduced. The long-term stability and cross-year consistency of the luminosity calibration are conclusively validated using 
the \PZ boson production rate in the dimuon final state.

The precision of the van der Meer calibration procedure has reached the sub-percent level target set forward by the CMS Collaboration for achieving the physics goals of the (High-Luminosity) LHC.
The extrapolation to standard physics data-taking conditions (luminosity integration) remains
the leading source of uncertainty and is expected to become even more challenging with the intensity increase foreseen in the HL-LHC era. Further improvements in the emittance scan analysis methodology, as well as in \PZ boson rate measurements, 
will play an important role in overcoming this limitation.

The 0.73\% relative precision achieved for the complete Run~2 proton-proton data set stands as the most precise luminosity measurement ever performed at a bunched-beam hadron collider. This milestone establishes a pivotal benchmark for the broader physics community, strengthening the impact of LHC data on precision measurements and constraints on new physics.

\begin{acknowledgments}
We congratulate our colleagues in the CERN accelerator departments for the excellent performance of the LHC and thank the technical and administrative staffs at CERN and at other CMS institutes for their contributions to the success of the CMS effort. In addition, we gratefully acknowledge the computing centers and personnel of the Worldwide LHC Computing Grid and other centers for delivering so effectively the computing infrastructure essential to our analyses. Finally, we acknowledge the enduring support for the construction and operation of the LHC, the CMS detector, and the supporting computing infrastructure provided by the following funding agencies: SC (Armenia), BMBWF and FWF (Austria); FNRS and FWO (Belgium); CNPq, CAPES, FAPERJ, FAPERGS, and FAPESP (Brazil); MES and BNSF (Bulgaria); CERN; CAS, MoST, and NSFC (China); MINCIENCIAS (Colombia); MSES and CSF (Croatia); RIF (Cyprus); SENESCYT (Ecuador); ERC PRG and PSG, TARISTU24-TK10 and MoER TK202 (Estonia); Academy of Finland, MEC, and HIP (Finland); CEA and CNRS/IN2P3 (France); SRNSF (Georgia); BMFTR, DFG, and HGF (Germany); GSRI (Greece); MATE and NKFIH (Hungary); DAE and DST (India); IPM (Iran); SFI (Ireland); INFN (Italy); MSIT and NRF (Republic of Korea); MES (Latvia); LMTLT (Lithuania); MOE and UM (Malaysia); BUAP, CINVESTAV, CONACYT, LNS, SEP, and UASLP-FAI (Mexico); MOS (Montenegro); MBIE (New Zealand); PAEC (Pakistan); MSHE, NSC, and NAWA (Poland); FCT (Portugal); MESTD (Serbia); MICIU/AEI and PCTI (Spain); MOSTR (Sri Lanka); Swiss Funding Agencies (Switzerland); MST (Taipei); MHESI (Thailand); TUBITAK and TENMAK (T\"{u}rkiye); NASU (Ukraine); STFC (United Kingdom); DOE and NSF (USA).

\begin{sloppy}
\setlength\emergencystretch{\hsize}
\hyphenation{Rachada-pisek} Individuals have received support from the Marie-Curie program and the European Research Council and Horizon 2020 Grant, contract Nos.\ 675440, 724704, 752730, 758316, 765710, 824093, 101115353, 101002207, 101001205, and COST Action CA16108 (European Union); the Leventis Foundation; the Alfred P.\ Sloan Foundation; the Alexander von Humboldt Foundation; the Science Committee, project no. 22rl-037 (Armenia); the Fonds pour la Formation \`a la Recherche dans l'Industrie et dans l'Agriculture (FRIA) and Fonds voor Wetenschappelijk Onderzoek contract No. 1228724N (Belgium); the Beijing Municipal Science \& Technology Commission, No. Z191100007219010, the Fundamental Research Funds for the Central Universities, the Ministry of Science and Technology of China under Grant No. 2023YFA1605804, the Natural Science Foundation of China under Grant No. 12535004, and USTC Research Funds of the Double First-Class Initiative No.\ YD2030002017 (China); the Ministry of Education, Youth and Sports (MEYS) of the Czech Republic; the Shota Rustaveli National Science Foundation (Georgia); the Deutsche Forschungsgemeinschaft (DFG), among others, under Germany's Excellence Strategy -- EXC 2121 ``Quantum Universe" -- 390833306, and under project number 400140256 - GRK2497; the Hellenic Foundation for Research and Innovation (HFRI), Project Number 2288 (Greece); the Hungarian Academy of Sciences, the New National Excellence Program - \'UNKP, the NKFIH research grants K 131991, K 138136, K 143460, K 143477, K 147557, K 146913, K 146914, K 147048, TKP2021-NKTA-64, and 2025-1.1.5-NEMZ\_KI-2025-00004, and MATE KKP and KKPCs Research Excellence and Flagship Research Groups grants (Hungary); the Council of Science and Industrial Research, India; ICSC -- National Research Center for High Performance Computing, Big Data and Quantum Computing, FAIR -- Future Artificial Intelligence Research, and CUP I53D23001070006 (Mission 4 Component 1), funded by the NextGenerationEU program, the Italian Ministry of University and Research (MUR) under Bando PRIN 2022 -- CUP I53C24002390006, PRIN PRIMULA 2022RBYK7T (Italy); the Latvian Council of Science; the Ministry of Science and Higher Education, project no. 2022/WK/14, and the National Science Center, contracts Opus 2021/41/B/ST2/01369, 2021/43/B/ST2/01552, 2023/49/B/ST2/03273, and the NAWA contract BPN/PPO/2021/1/00011 (Poland); the Funda\c{c}\~ao para a Ci\^encia e a Tecnologia (Portugal); the National Priorities Research Program by Qatar National Research Fund;  MICIU/AEI/10.13039/501100011033, ERDF/EU, ``European Union NextGenerationEU/PRTR", projects PID2022-142604OB-C21, PID2022-139519OB-C21, PID2023-147706NB-I00, PID2023-148896NB-I00, PID2023-146983NB-I00, PID2023-147115NB-I00, PID2023-148418NB-C41, PID2023-148418NB-C42, PID2023-148418NB-C43, PID2023-148418NB-C44, PID2024-158190NB-C22, RYC2021-033305-I, RYC2024-048719-I, CNS2023-144781, CNS2024-154769 and Plan de Ciencia, Tecnolog{\'i}a e Innovaci{\'o}n de Asturias, Spain; the Chulalongkorn Academic into Its 2nd Century Project Advancement Project, the National Science, Research and Innovation Fund program IND\_FF\_68\_369\_2300\_097, and the Program Management Unit for Human Resources \& Institutional Development, Research and Innovation, grant B39G680009 (Thailand); the Eric \& Wendy Schmidt Fund for Strategic Innovation through the CERN Next Generation Triggers project under grant agreement number SIF-2023-004; the Kavli Foundation; the Nvidia Corporation; the SuperMicro Corporation; the Welch Foundation, contract C-1845; and the Weston Havens Foundation (USA).
\end{sloppy}
\end{acknowledgments}\section*{Data availability} Release and preservation of data used by the CMS Collaboration as the basis for publications is guided by the  \href{https://doi.org/10.7483/OPENDATA.CMS.1BNU.8V1W}{CMS data preservation, re-use and open access policy}.

\bibliography{auto_generated}
\cleardoublepage \appendix\section{The CMS Collaboration \label{app:collab}}\begin{sloppypar}\hyphenpenalty=5000\widowpenalty=500\clubpenalty=5000\input{LUM-20-001-public-authorlist.tex}\end{sloppypar}
%%% END EDITABLE REGION %%%
% skeleton_end
\end{document}

%% file: LUM-20-001-public-authorlist.tex
\cmsinstitute{Yerevan Physics Institute, Yerevan, Armenia}
{\tolerance=6000
A.~Gevorgyan\cmsorcid{0000-0003-2751-9489}, A.~Hayrapetyan, V.~Makarenko\cmsorcid{0000-0002-8406-8605}, A.~Tumasyan\cmsAuthorMark{1}\cmsorcid{0009-0000-0684-6742}
\par}
\cmsinstitute{Marietta Blau Institute for Particle Physics, Vienna, Austria}
{\tolerance=6000
W.~Adam\cmsorcid{0000-0001-9099-4341}, L.~Benato\cmsorcid{0000-0001-5135-7489}, T.~Bergauer\cmsorcid{0000-0002-5786-0293}, M.~Dragicevic\cmsorcid{0000-0003-1967-6783}, S.~Gundacker\cmsorcid{0000-0003-2087-3266}, A.K.~Guven\cmsorcid{0009-0004-5670-5138}, P.S.~Hussain\cmsorcid{0000-0002-4825-5278}, M.~Jeitler\cmsAuthorMark{2}\cmsorcid{0000-0002-5141-9560}, N.~Krammer\cmsorcid{0000-0002-0548-0985}, A.~Li\cmsorcid{0000-0002-4547-116X}, D.~Liko\cmsorcid{0000-0002-3380-473X}, M.~Matthewman, J.~Schieck\cmsAuthorMark{2}\cmsorcid{0000-0002-1058-8093}, R.~Sch\"{o}fbeck\cmsAuthorMark{2}\cmsorcid{0000-0002-2332-8784}, M.~Shooshtari\cmsorcid{0009-0004-8882-4887}, M.~Sonawane\cmsorcid{0000-0003-0510-7010}, N.~Van~Den~Bossche\cmsorcid{0000-0003-2973-4991}, W.~Waltenberger\cmsorcid{0000-0002-6215-7228}, C.E.~Wulz\cmsAuthorMark{2}\cmsorcid{0000-0001-9226-5812}
% ...
\par}
$^{1}$Also at Yerevan State University, Yerevan, Armenia\\
$^{2}$Also at Technische Universit\"{a}t Wien, Vienna, Austria\\
% ...
